%% file: paper.tex
\documentclass[acm,sigplan]{acmart}

\input{packages}
\input{preamble}

\begin{document}
\hypersetup{
  linkcolor=purple,
  citecolor=blue,
}

\newcommand{\hd}[1]{\noindent{\textcolor{purple}{\textbf{\fbox{hd} \textit{#1}}}}}

\newcommand{\yc}[1]{\noindent{\textcolor{red}{\bf \fbox{yc} {\it#1}}}}


\title{\LARGE \bf gigiProfiler: Diagnosing Performance Issues by Uncovering
\\ Application Resource Bottlenecks}

\author{Yigong Hu}
\affiliation{
  \institution{Boston University and University of Washington}
  \city{}
  \country{}}

\author{Haodong Zheng}
\affiliation{
  \institution{University of Washington}
  \city{}
  \country{}}

\author{Yicheng Liu}
\affiliation{
  \institution{University of California, Los Angeles}
  \city{}
  \country{}}

\author{Dedong Xie}
\affiliation{
  \institution{University of Washington}
  \city{}
  \country{}}

\author{Youliang Huang}
\affiliation{
  \institution{Boston University}
  \city{}
  \country{}}

\author{Baris Kasikci}
\affiliation{
  \institution{University of Washington}
  \city{}
  \country{}}

\settopmatter{printfolios=true,printacmref=false}

\maketitle 
\pagestyle{plain}

\input{section/abstract}

\input{section/intro}

\input{section/motivation}

\input{section/identification}

\input{section/design}

\input{section/eval}

\input{section/related}
\input{section/conclusion}

\clearpage
\bibliographystyle{ACM-Reference-Format}
\bibliography{bib/references}

\end{document}

%% file: packages.tex
\usepackage{epsfig}
\usepackage[T1]{fontenc}
\usepackage{newtxtext}
\usepackage[scaled=0.75]{beramono}
\usepackage{newtxmath}
\usepackage[ruled,noend,linesnumbered,vlined]{algorithm2e}
\usepackage{listings}
\usepackage{xcolor}
\usepackage{colortbl}
\usepackage{balance}
\usepackage{graphicx}
\usepackage{subcaption}
\usepackage{wrapfig}
\usepackage[font={small},labelfont=bf]{caption}
\usepackage{booktabs}
\usepackage[export]{adjustbox}
\usepackage{datetime}
\usepackage{enumitem}
\usepackage{multirow}
\usepackage[frozencache,cachedir=.]{minted}
\usepackage{siunitx}
\usepackage{pifont}
\usepackage{tikz}
\usetikzlibrary{tikzmark,calc}
\usepackage{cleveref}
\usepackage{amsmath}
\usepackage{makecell}
\usepackage{algorithmic}

\usepackage{scalefnt}

%% file: preamble.tex
\graphicspath{{./}{figure/}}
\usepackage{url}

\setminted{style=tango,breaklines=true,breaksymbolleft=,fontsize=\small}
\DeclareMathAlphabet{\mathcal}{OMS}{cmsy}{m}{n}

\newcommand{\tool}{\textit{\mbox{gigiProfiler}}\xspace}
\newcommand{\trace}{\textit{\mbox{OmniResource Profiling}}\xspace}

\DeclareSIUnit{\operation}{op}
\definecolor{ngray}{RGB}{102,102,102}
\renewcommand\footnotetextcopyrightpermission[1]{}

\newcolumntype{R}[2]{%
  >{\adjustbox{angle=#1,lap=\width-(#2)}\bgroup}%
  l%
  <{\egroup}%
}

\definecolor{bg}{rgb}{0.95,0.95,0.95}

\crefname{section}{\S}{\S\S}
\crefname{subsection}{\S}{\S\S}
\crefformat{section}{\S#2#1#3}
\crefformat{subsection}{\S#2#1#3}

\newcounter{magicrownumbers}
\newcommand\rownumber{\stepcounter{magicrownumbers}\arabic{magicrownumbers}}

\makeatletter
\patchcmd{\@algocf@start}
  {-1.5em}
  {0pt}
  {}{}
\makeatother

%% file: section/abstract.tex
\section*{Abstract}

Diagnosing performance bottlenecks in modern software is essential yet 
challenging, particularly as applications become more complex and rely on custom resource 
management policies. While traditional profilers effectively identify execution 
bottlenecks by tracing system-level metrics, they fall short when it comes to 
application-level resource contention caused by waiting for application-level events.

In this work, we introduce \trace, a performance analysis approach that integrates system-level and application-level resource tracing to diagnose resource bottlenecks comprehensively. \tool, our realization of \trace, uses a hybrid LLM-static analysis approach to identify application-defined resources offline and analyze their impact on performance during buggy executions to uncover the performance bottleneck. \tool then samples and records critical variables related to these bottleneck resources during buggy execution and compares their value with those from normal executions to identify the root causes. We evaluated \tool on 12 real-world performance issues across five applications. \tool accurately identified performance bottlenecks in all cases. \tool also successfully diagnosed the root causes of two newly emerged, previously undiagnosed problems, with the findings confirmed by developers.

%% file: section/intro.tex
\section{Introduction}
\label{sec:intro}

Diagnosing performance issues in modern software is both essential and challenging, particularly as applications grow more complex. Among various performance issues, resource bottlenecks, caused by poor resource management, inefficient algorithms, or contention between application components, are especially difficult to diagnose. Identifying these bottlenecks requires understanding not only where resources are consumed but also how contention arises from interactions within the application. Developers often turn to state-of-the-art profilers~\cite{ren2019relational,yang2016wperf,Weng2023vProf,curtsinger2015coz,attariyan2010xray} to analyze execution time and locate potential bottlenecks. For example, CoZ~\cite{curtsinger2015coz} identifies optimization opportunities by analyzing causal relationships in program execution, while relational debugging~\cite{ren2019relational} captures and reasons about the causal relationships between input and output events to diagnose performance issues.

However, these tools often fail to capture the interactions within applications that are critical to diagnosing performance issues. Many performance issues are caused by application-defined resource contention~\cite{mysql_bug_75540,percona_swapping,mysql_bug_99315} such as buffer pool evictions or log contention, which are often overlooked. Previous studies~\cite{percona_hung_transaction,hu23pbox,yhuelf_pg_stat_statements,percona_xtrabackup} have shown that these interactions are essential for understanding the root causes of performance degradation. Without a tool that captures application-level resource interactions, developers are forced to rely on their own understanding of the software or manually add instrumentation to trace resource usage and contention, making the process both time-consuming and error-prone.

\begin{figure}[t]
    \centering
    \begin{subfigure}[t]{0.45\textwidth}
        \centering
        \includegraphics[width=\textwidth]{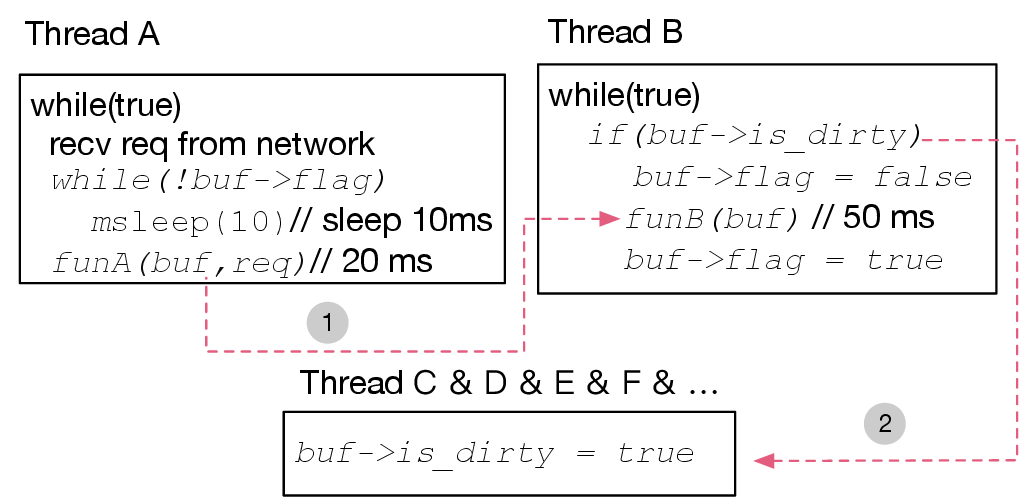}
        \caption{Pseudo code of a resource contention issue in MySQL.}
        \label{fig:example_a}
    \end{subfigure}
    \hfill
    \begin{subfigure}[t]{0.45\textwidth}
        \centering
        \includegraphics[width=\textwidth]{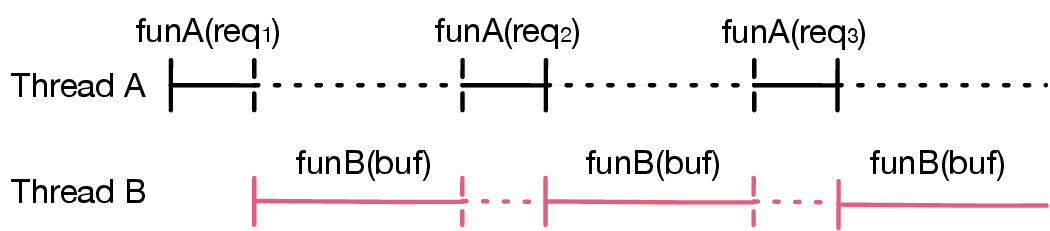} 
        \caption{OmniResource profiling to diagnose the contention.}
        \label{fig:example_b}
    \end{subfigure}
    \caption{(a) Example of resource contention in program logic. (b) Diagnosing the contention using OmniResource Profiling.}
    \label{fig:example}
\end{figure}

Figure~\ref{fig:example_a} shows the pseudocode of a real-world performance issue~\cite{percona_swapping} in MySQL. When the \textit{buf} is dirty, thread B must call \textit{funcB} to clean it, a process that takes approximately 50 ms. During this cleanup, thread A, which is processing a request, must wait for the \textit{buf} to become available. Thread A periodically checks the flag and yields execution until the cleanup is complete. As a result, the request in thread A takes 70 ms to finish instead of the expected 20 ms. The root cause of this issue is frequent writes by other threads to the \textit{buf}, marking it as dirty and creating buffer pool contention.

To identify the root cause of this performance issue, developers must locate the buffer  contention between thread A and thread B, as well as determine which threads are marking \textit{buf} as dirty. Unfortunately, existing tools are limited to reporting system-level waiting events (e.g., lock or I/O wait times). However, in this case, there are no explicit system-level waiting events—thread A simply sleeps while repeatedly checking the \textit{is\_dirty} flag, and the contention arises entirely from application-level logic.  While manually identifying resource contention in a simple example like Figure~\ref{fig:example} may be manageable, it becomes challenging for developers to locate the root cause in large-scale, complex software.

This paper proposes \trace, which integrates system-level and application-level resource tracing to diagnose application resource bottlenecks. The key idea is to trace application resources with the same level of detail as system resources to identify bottlenecks comprehensively. Figure~\ref{fig:example_b} shows how \trace is used to diagnose the root cause of contention on \textit{buf}. \trace first identifies \textit{buf} as application resource and traces its usage. It records that the thread A is waiting on thread B and the time spent in thread B is in \textit{funB}.

A core challenge in enabling \trace is accurately identifying application-defined resources and their usage. These resources are often deeply embedded within the internal logic of the application and are not directly observable by external monitoring tools. Moreover, their design and implementation are tightly coupled with application-specific semantics and developer preferences, leading to a lack of common code patterns. Consequently, traditional static analysis techniques face significant challenges in accurately and efficiently detecting these resources. Static analysis must either extend its coverage to include a wide range of application-level code patterns, risking false positives, or limit its scope to specific patterns and rely on manual annotations, which reduces coverage. Both approaches significantly limit their effectiveness in diagnosing performance issues.

Our key insight is that application-defined resources' information can be inferred from program metadata, such as function definitions, comments, and documentation. Based on this, we propose a hybrid approach that combines large language models (LLMs) with static analysis to accurately identify application-defined resources and their usage. LLMs analyze metadata to infer high-level application semantics, enabling broad coverage in identifying potential resources and usage. Static analyzer then validates these inferred results, ensuring accuracy by leveraging precise code patterns. By dividing the task into two steps—LLMs for high coverage and static analysis for validation—the approach overcomes the traditional trade-off static analysis faces between coverage and accuracy. This combination allows the hybrid method to achieve comprehensive resource discovery and precise identification, making it particularly effective for diagnosing performance issues.

To show the effectiveness of \trace, we developed a tool called \tool. Using the hybrid LLM-static analysis approach described earlier, \tool identifies application-defined resources and their operator functions offline. After this step, \tool runs a buggy execution to uncover performance bottlenecks and associates them with the identified resources. Beyond identifying bottleneck resources, \tool seeks to explain why the contention occurs in the code. To achieve this, \tool records the values of critical variables related to these resources during the buggy execution. These recorded values are then compared with those from a normal execution to identify anomalies. For example, to understand why \textit{buf} is frequently marked as dirty, \tool samples and records the value of \textit{buf->dirty} during buggy and normal executions. By comparing these samples, \trace reveals that in the buggy execution, the value of \textit{buf->dirty} is frequently modified by specific client threads absent in the normal execution. 

A key challenge for \tool is minimizing the overhead associated with recording variable values. Capturing all program variables and their complete data flow would be prohibitively expensive, making \trace impractical for large-scale applications. To address this, \tool uses static analysis to identify specific program variables that influence application-defined resource usage. During execution, \tool selectively samples and records the values of only these relevant variables, significantly reducing overhead. After profiling, \tool compares the recorded variable values from the buggy execution with those from the normal execution to identify anomalous values that contribute to performance bottlenecks. 

We applied \tool to five large-scale real-world applications—MySQL, MariaDB, PostgreSQL, Apache, and LLAMA—and demonstrated its effectiveness in diagnosing 12 real-world performance issues. Our evaluation shows that \tool can accurately identify resource bottlenecks and pinpoint critical root-cause variables, providing valuable insights into the underlying causes of performance issues. In contrast, existing tools are limited to identifying system-level waiting events and fail to provide the contextual information necessary to diagnose these events effectively. Notably, \tool was applied to two unresolved issues where developers had been unable to locate the root causes; \tool successfully identified the root causes, which were later confirmed by the developers.

%% file: section/motivation.tex
\section{\trace}
\label{sec:example}

This section presents a motivating example to demonstrate the effectiveness of \trace in diagnosing complex performance issues within applications. The example focuses on an UNDO log contention problem in MySQL, which has been a major source of performance regressions in real-world deployments over the past decade~\cite{alibaba_undo_logs,percona_multiversioning,percona_transaction_history,percona_hung_transaction,mysql_bug_75540,percona_isolation_modes}.

\subsection{Motivating Example: UNDO Log Contention}
\label{sec:example_undo_log}

The UNDO log is a data structure used in databases to maintain consistency by tracking changes made during a transaction, allowing the system to roll back to a previous state if needed. MySQL implements the UNDO log as a linked list, where each node records the difference between two table versions before and after a transaction modifies the table. When a transaction is committed, the purge thread locks the UNDO log to remove old records associated with that transaction, blocking queries that attempt to write to the same table until the cleanup process is complete. Since accessing an older version requires iterating through the linked list, the UNDO log becomes a bottleneck resource, and its use by the purge thread can significantly delay other queries.

\begin{figure}[t]
  \centering
    \includegraphics[width=3in]{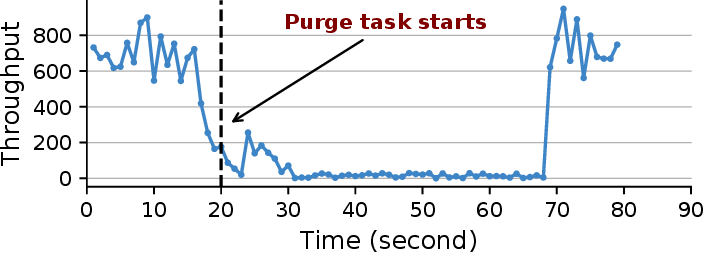}
    \caption{Throughput of all clients, highlighting a significant drop in throughput caused by the purge thread.}
    \label{fig:study_undolog}
\end{figure}

To reproduce contention on the UNDO log, we created a table with 1,000 records and ran two clients, A and B. Client A executed multiple read queries in a transaction and committed after 20 seconds. Concurrently, Client B issued write queries on the same table, resulting in the creation of a long UNDO log. Once Client A’s transaction was committed, the purge threads started cleaning the UNDO log, making Client B’s throughput plummet almost to zero for 10 seconds after the purge began. The throughput remained low until the purge threads finished. As shown in Figure~\ref{fig:study_undolog}, this delay caused Client B to take 80 seconds to complete all its queries—twice as long as the expected execution time.

\begin{figure}[t]
  \centering
    \includegraphics[width=3in]{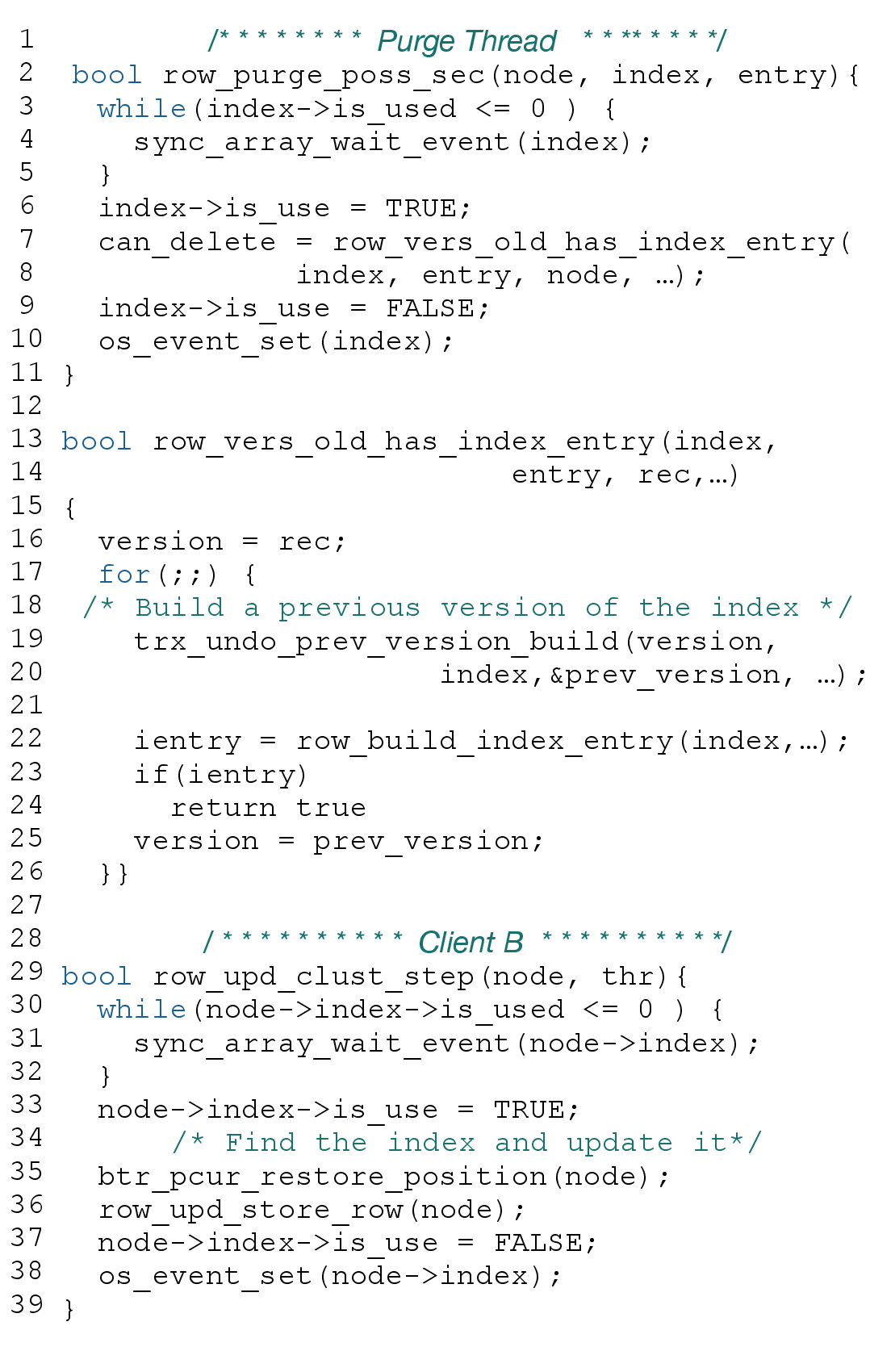}
    \caption{Code snippets showing how the UNDO log contention cause performance issues in Client B.}
    \label{fig:code_undolog}
\end{figure}

\begin{figure}[t]
  \centering
    \includegraphics[width=3in]{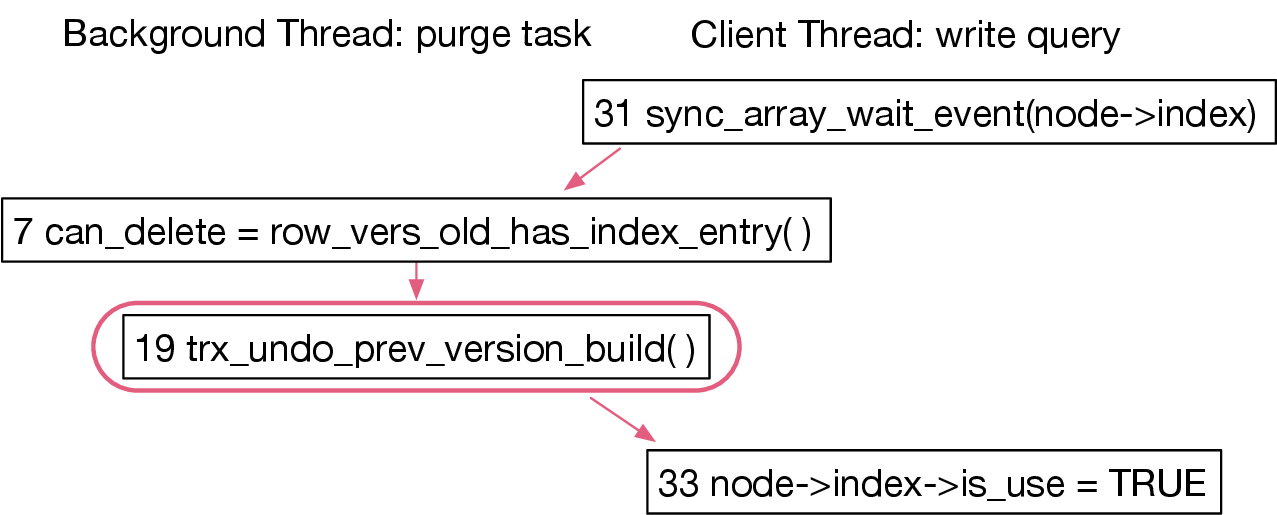}
    \caption{The root cause of UNDO log case inferred by \trace.}
    \label{fig:root_cause}
\end{figure}

Figure~\ref{fig:code_undolog} provides a simplified code snippet illustrating the root cause of the UNDO log contention. The contention occurs at line 31, where Client B’s update query waits for the purge thread to release \texttt{node->index}, which represents the UNDO log. The purge thread locks \texttt{node->index} at line 6 while checking if an old snapshot can be safely deleted. This involves calling \texttt{trx\_undo\_prev\_version\_build}, which iterates through the undo log’s linked list to incrementally construct each version until reaching the target. For large undo logs, this process can take 30 to 100 milliseconds, resulting in prolonged locking of \texttt{node->index}. This delays Client B’s update queries, making the UNDO log a bottleneck and severely impacting query throughput.

\subsection{\trace to Diagnose the Root Cause}

Existing performance debugging techniques struggle with this type of issue. First, this case is not caused by a slow execution path, so identifying performance hotspots (e.g., functions consuming high CPU time) offers no insight. Second, the UNDO log contention is not managed by system-level mechanisms like mutexes or system calls, making system resource tracking tools like \textit{perf} ineffective. Third, statistical debugging also fails, as the issue lacks observable anomalies, such as unexpected branches or unusual return values.

As a result, diagnosing such performance issues often requires significant manual effort from developers. While the example in Section~\ref{sec:example_undo_log} seems straightforward after identifying the root cause, the real challenge lies in uncovering contention that happened in complex interactions between the purge thread and multiple client queries, hidden within application-specific logic. Developers must pinpoint the UNDO log as a potential bottleneck, instrument the code to trace its usage, and analyze interactions between components. This process is time-consuming, error-prone, and demands deep expertise in the application’s internal workings.

The previous example shows that the primary challenge in diagnosing this performance issue is that the bottleneck resource, UNDO log, is an application-defined resource with custom usage policies that are invisible to system-level tools. If we could trace application-defined resources with the same granularity as system resources, identifying the root cause would become significantly more straightforward.

\trace tackles this challenge by combining system-level and application-level resource tracing to diagnose performance bottlenecks. As shown in Figure~\ref{fig:root_cause}, it identifies \texttt{node->index} as an application-defined resource and the key functions in the purge thread. By tracing and analyzing the purge thread’s cleanup time and the client thread’s waiting time, \trace pinpoints the UNDO log as the bottleneck and \texttt{trx\_undo\_prev\_version\_build} as the function causing prolonged holding of the resource.


\subsection{Opportunities of Enabling \trace}

Enabling OmniResource Profiling presents several challenges. First, application resources are often deeply integrated into application logic, making them invisible to external monitoring tools. For example, MySQL implements custom event-waiting functions for the UNDO log rather than using standard mutex system calls. Second, profiling all synchronization-related system calls indiscriminately creates excessive overhead and irrelevant data, such as tracing the locking time of a short operation like updating a global variable. Thus, OmniResource Profiling must precisely identify and track application-defined resources to minimize overhead.

Identifying application-defined resources and their usage is particularly challenging due to the diversity in how they are implemented across applications. For example, MySQL represents the UNDO log as a linked list of undo log records, whereas PostgreSQL implements it as redundant rows in a table with older transaction IDs. These application-specific differences make it difficult to generalize code patterns for detecting application resources. Mapping each resource to its specific code pattern requires substantial domain knowledge and is highly error-prone.

As a result, existing static analysis techniques often rely on manual annotations to guide the analysis. This not only limits their effectiveness in diagnosing performance issues but also restricts their applicability to broader scenarios. To the best of our knowledge, no existing tools can automatically identify application-defined resources to enable profiling.

Despite these challenges, we observe that most software has abundant documentation and comments, which we refer to as metadata in this paper. These metadata provide valuable insights into the application’s design and implementation, such as the purpose of functions, relationships between modules, and, importantly, the definitions of application-defined resources. Developers can often use this metadata to identify potential application resources. For example, the function header comment of \texttt{trx\_undo\_prev\_version\_build} contains the description: \textit{Build a previous version of a clustered index record. The caller must hold a latch on the index page of the clustered index record.} This description implies that \texttt{node->index} is a key resource shared by multiple threads, making it a potential bottleneck in certain scenarios.

Based on this observation, we propose leveraging metadata to infer high-level application semantics and identify potential application-defined resources. Large Language Models have demonstrated exceptional capabilities in processing such metadata~\cite{Chen2024rcacopilot,xie2024resym,Ahmed2023rootcause}, often rivaling human experts in code understanding tasks. By applying LLMs to analyze software documentation, comments, and even source code, we can systematically uncover application-defined resources and their semantics, creating new opportunities to diagnose performance bottlenecks effectively.

Unlike traditional approaches that rely on predefined rules or manual annotations, LLMs offer strong generality, allowing them to adapt to a wide variety of applications without requiring extensive customization. Additionally, LLMs excel at handling ambiguity in metadata, such as incomplete or noisy documentation and comments, by leveraging their pretrained knowledge and contextual understanding. These capabilities make LLMs particularly effective in inferring application-defined resource accesses in diverse applications.


%% file: section/identification.tex
\section{Identifying Application-Defined Resources}
\label{sec:design}

In this section, we describe how we identify application-defined resources to enable \trace. We propose a hybrid analysis approach that combines LLMs with static analysis. The LLM module analyzes application documentation to infer potential candidate resources. A static analysis then validates these inferred resources based on the code pattern to ensure the LLM’s output is both accurate.

\subsection{Lesson from Our Initial Attempt}
\label{sec:design_opportunities}

We began by developing a prompt-based LLM module to identify application-defined resources and their operator without incorporating static analysis. The prompts provided a definition of the resources and functions that operate it, and a concrete example to guide the LLM’s inference. However, our initial attempts at using LLMs for this sophisticated task revealed significant limitations. While LLMs could identify many application-defined resources, their accuracy was inconsistent. Simply relying on prompts to enhance the LLM’s understanding of application-defined resources resulted in poor overall performance. Our study identified four key challenges that contribute to LLMs’ misinterpretation.

\textbf{Missing software context}: Application documentation often lacks critical details like control flow and data flow, which are necessary for the LLM to infer how resources are used within a function and determine whether the function contributes to resource contention or plays a peripheral role. For example, while the comments of \texttt{binlog\_init} suggest it uses the binlog buffer (an application resource), this function is called only once during initialization and should not be flagged as a significant operator for the resource.

\textbf{Misleading descriptions}: Ambiguous function comments can mislead LLMs. For example, the function \texttt{is\_active} merely checks the status of a log file and does not interact with shared resources. However, its description includes information suggesting that the log is a critical resource used in this function, causing the LLM to mistakenly mark it as a key operator function on the log resource.

\textbf{Comment loss}: The absence of comments or other contextual information leaves the LLM with too little data to accurately interpret a function’s role. For example, the function \texttt{is\_number} has no comment, providing the LLM with only the function name as input. As a result, while this function parses strings of binary logs and serves as a key operator for the log resource, the LLM fails to recognize its importance based solely on its name.

\textbf{Misinterpretation}: Even when function comments are clear, the LLM can generate incorrect insights due to the inherent complexity of the application logic.

Besides these challenges, the LLM’s output also suffers from inconsistency due to the probabilistic nature of language models. Even for the same function with the same prompt, the LLM can generate different results in multiple runs. This inconsistency poses significant challenges for identifying resources systematically and meeting the consistency requirements of performance debugging.

Our lessons from these limitations reveal that LLMs alone are insufficient for accurately identifying application-defined resources. Even with improved prompts, some challenges such as \textit{missing software context} are fundamentally hard to address without the help of static analysis. Also, comments or source code alone may not provide enough detail to infer the usage patterns of shared resources. These observations motivated the development of our hybrid approach.

\subsection{Our Hybrid Approach}

Our initial attempt suggests that treating the identification of application-defined resources as a single, monolithic task inevitably results in trade-offs between coverage and accuracy. Neither LLMs nor static analysis can simultaneously achieve both high coverage and accuracy, as these goals often conflict.

\begin{figure}[t]
    \centering
      \includegraphics[width=3in]{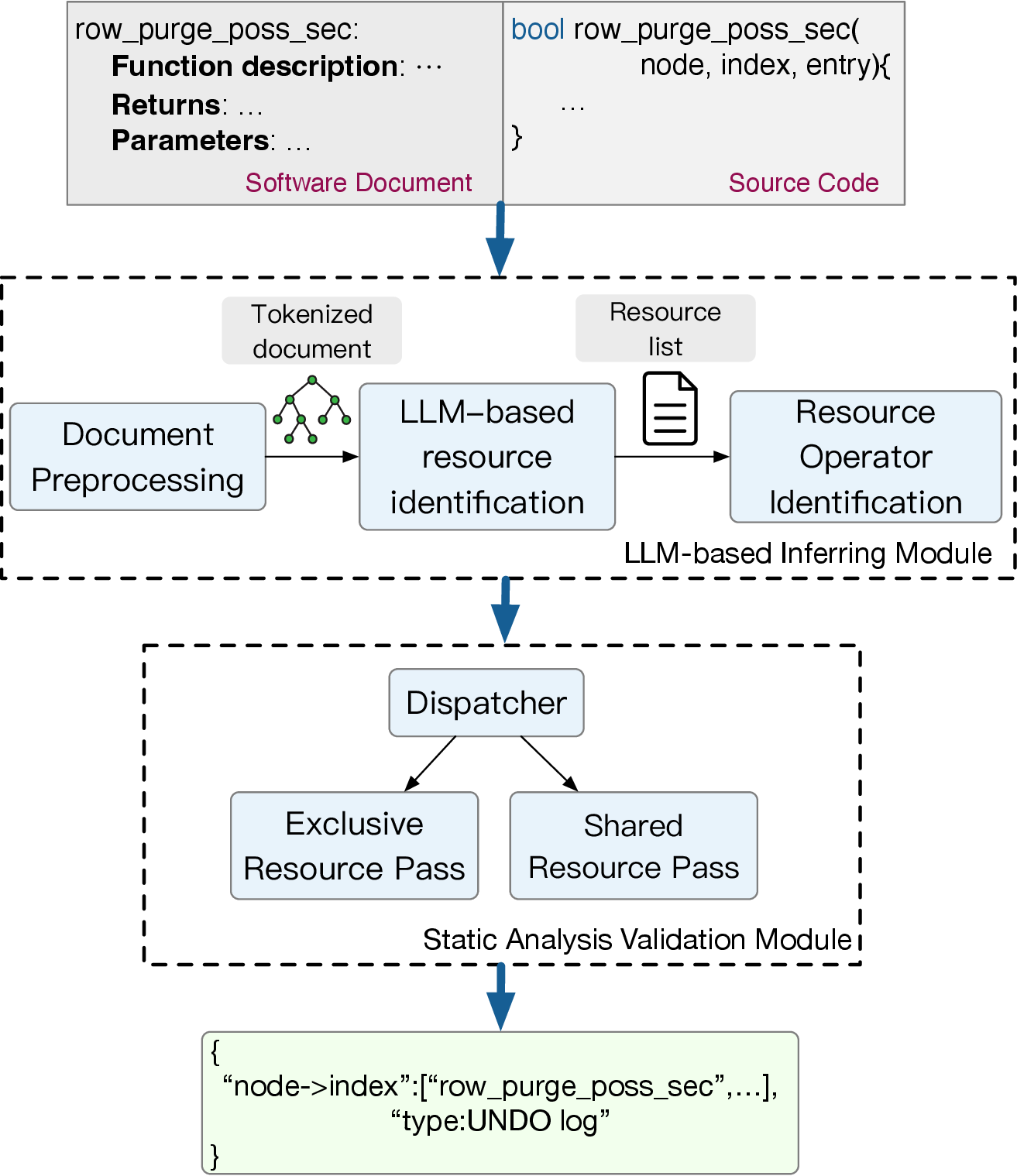}
      \caption{Overview of the hybrid approach.}
      \label{fig:overview_hybrid}
  \end{figure}

Our key insight is to divide the problem into two complementary subtasks: detecting candidate resources based on application metadata and validating these candidates using concrete code patterns. This decomposition leverages the strengths of each approach. LLM module can focus on achieving high coverage by analyzing high-level semantics from documentation and comments, while the static analyzer ensures accuracy by validating results against code patterns. By assigning each subtask to the method best suited for it, the hybrid approach achieves both high coverage and accuracy in the identification of application-defined resources.

An additional benefit of this hybrid approach is that the static analyzer uses code patterns for final validation, ensuring the results are explainable. This explainability is essential for developers, as it provides clarity about the identified resources and helps them understand the root causes of performance issues, fostering trust in the tool’s output.

Figure~\ref{fig:overview_hybrid} shows the workflow of our hybrid approach. The process begins with document preprocessing, where software metadata is tokenized and organized into a tree hierarchy for structured analysis. The preprocessed metadata is then analyzed by the LLM in two stages. Resource identification, where the LLM analyzes documentation, comments, and descriptions to infer high-level semantics and generate a list of candidate resources. The operator function identification stage then examines metadata and source code to associate these resources with potential operator functions.

The output from the LLM module is then fed into the static analysis validation module. This module employs a dispatcher to classify resources as either exclusive or shared. Each resource type undergoes validation through its respective analysis pass. The output of the hybrid approach is a validated resource list, detailing resource types, associated functions, and their usage patterns, ensuring both accuracy and comprehensiveness.

\subsection{LLM Module to Infer Resource Candidates}
\label{sec:llm_module}

The goal of the LLM module is to achieve high coverage in identifying application-defined resources while maintaining acceptable accuracy. A straightforward approach would be to use a single, static prompt to query the LLM for application resources. However, large-scale software like MySQL typically contains tens of application-defined resources, each with numerous usage points, spread across thousands of files in documentation and comments. This scale and complexity, combined with the inherent ambiguity in software documentation, make it impractical for a single prompt to reliably uncover all relevant resources and usage points. A more guided approach is required to help the LLM systematically navigate and analyze this extensive and complex information.

Inspired by human experts who iteratively examine documentation, comments, and code structure, we employ the Chain-of-Thought(CoT) prompting to guide the LLM in identifying application-defined resources and their usage points. The COT prompting~\cite{wei2024chainofthought} enhances the reasoning capabilities of LLMs by incorporating logical steps, or a “chain of thought”, within the prompt. Unlike direct-answer prompting, CoT guides the model to work through intermediate reasoning steps, making it more adept at solving complex tasks. 

We divide the task of identifying application-defined resources into two subtasks: (1) identifying potential resources and (2) identifying the operator functions for these resources. For each subtask, we design a series of CoT-based prompts to systematically guide the LLM in extracting relevant information from the application’s documentation and comments. To realize this process, we implement a three-step approach in our LLM module to identify resource candidates and their usage points effectively.

\begin{figure}[t]
    \centering
      \includegraphics[width=2.8in]{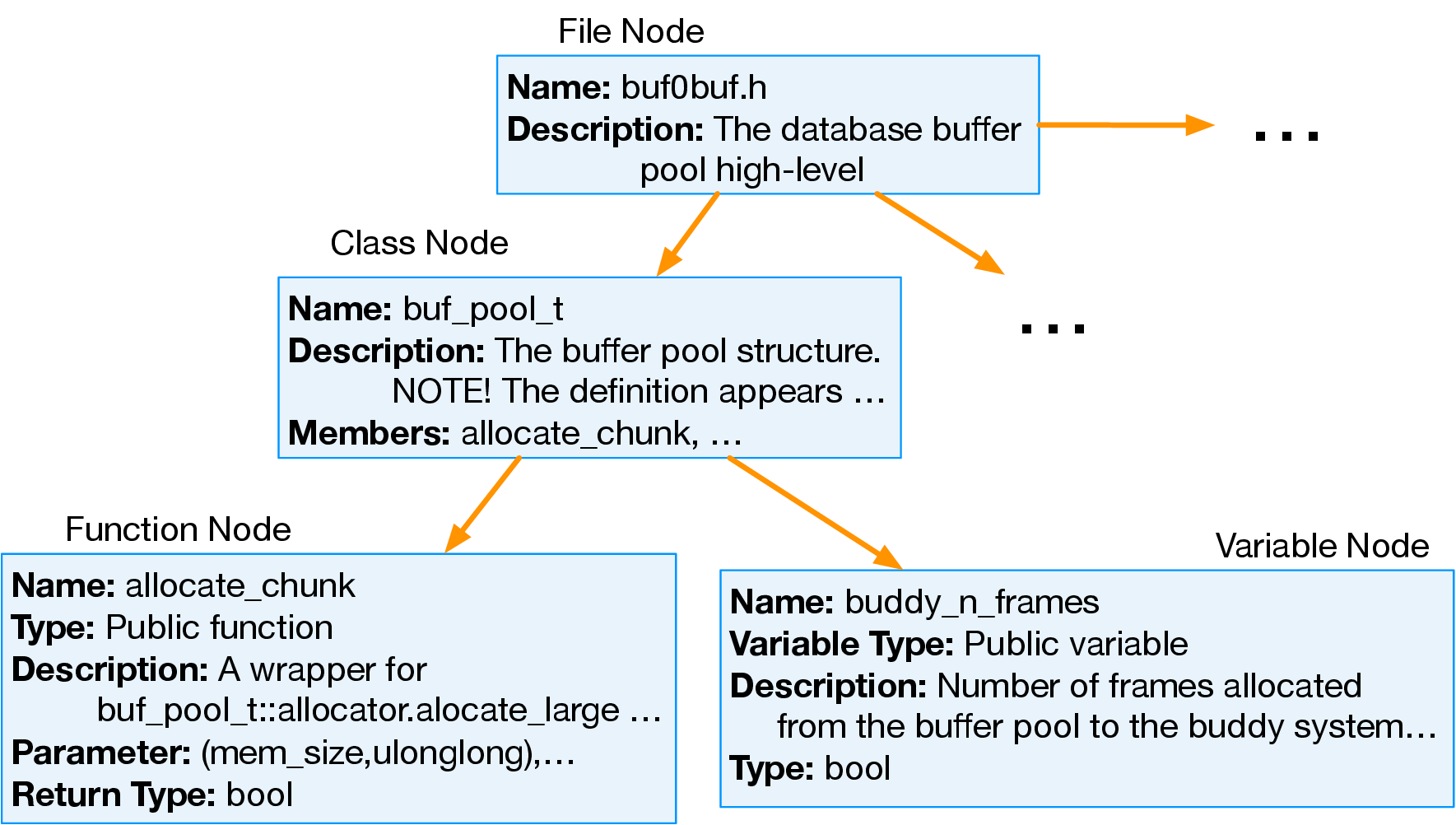}
      \caption{The tokenization format.}
      \label{fig:token_format}
      \vspace{-1em}
  \end{figure}

\textbf{Step 1:preprocessing software metadata.}
LLMs face input size limits, requiring software metadata to be processed in smaller batches. This division often leads to a loss of contextual information between functions, classes, and other elements of the codebase, increasing the risk of inaccuracies in analysis. To address this, we preprocess the metadata into a compressed and organized format. By structuring the data hierarchically, we can include more related information within a single batch, thereby reducing misinterpretation and improving the LLM’s overall accuracy and coverage.

Our tokenizer organizes software metadata into a tree hierarchy, with files as the top-level nodes containing classes, functions, and global variables as child nodes. Figure~\ref{fig:token_format} illustrates the tokenized format for a buffer pool file. Each level stores relevant details: files record filenames and file descriptions; classes store their descriptions, member functions, and variables; functions include details such as descriptions, parameters, and return values; variables specify their types, descriptions, and names. By organizing metadata in this way, we maintain contextual relationships and enable the LLM to process the data in a structured and coherent manner.

These metadata objects are stored in memory and can be easily saved and restored using Python’s \texttt{pickle} package. This approach allows for flexible manipulation of the metadata, enabling the generation of content at varying levels of detail to support CoT-based analysis during prompt preparation. Additionally, the objects can be formatted to meet the specific input requirements of the LLM, ensuring compatibility and ease of integration.

\begin{figure}[t]
    \centering
      \includegraphics[width=3in]{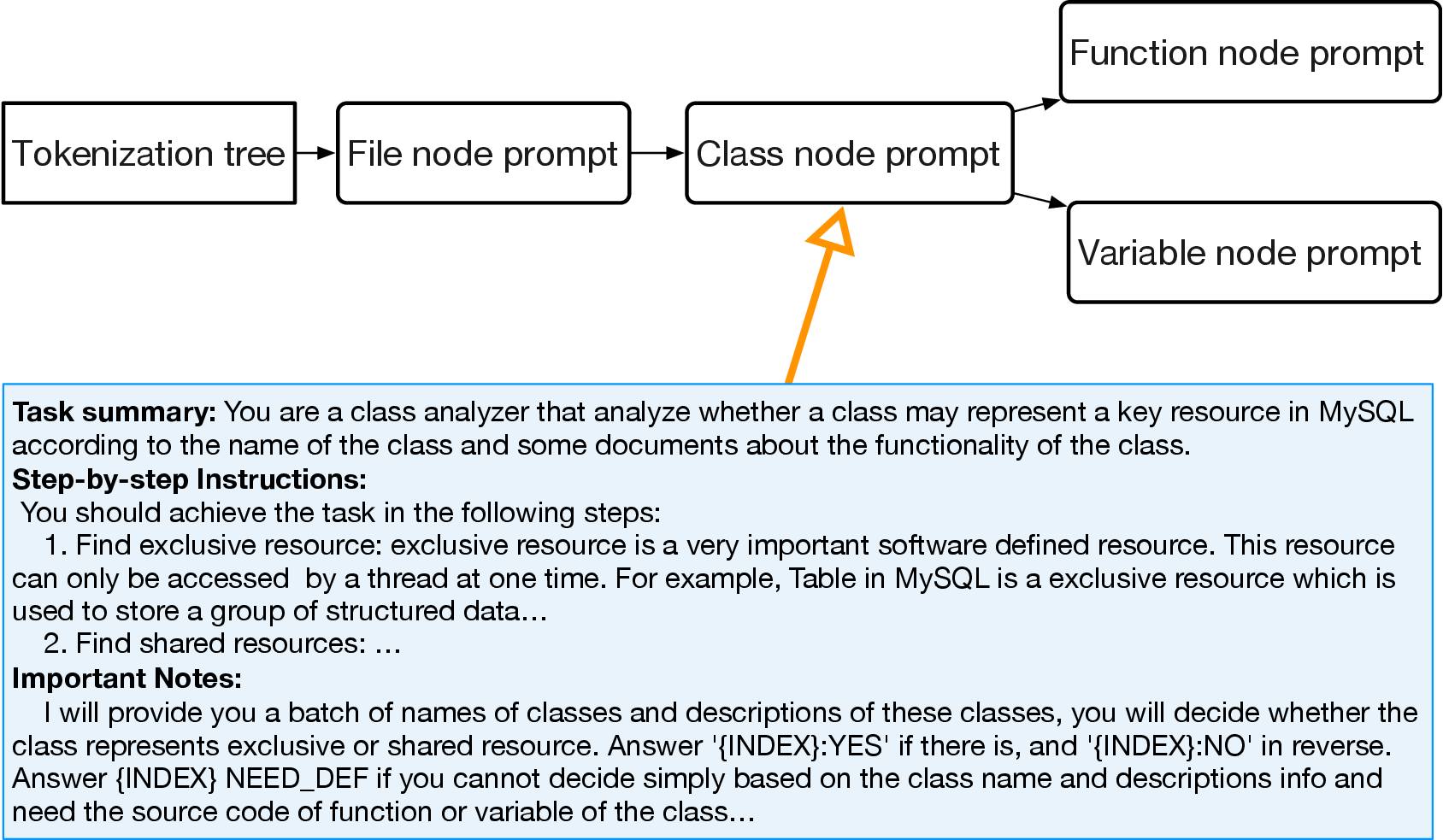}
      \caption{The workflow and prompts snapshot for step 2}
      \label{fig:step2_prompt}
      \vspace{-1em}
  \end{figure}
  
\textbf{Step 2: identifying potential resources.} 
Even with the advanced capabilities of LLMs, identifying application-defined resources requires carefully designed prompts to guide the model. Unlike system-level resources with standardized definitions, application-defined resources exhibit diverse usage patterns, making it challenging to provide a universal definition. Based on our study, application-defined resources can be categorized into two types: \textit{exclusive resources}, which are accessible by only one thread at a time, such as UNDO logs, and \textit{shared resources}, which are concurrently accessed by multiple threads, such as buffer pools and queues. This categorization enables us to design prompts that systematically guide the LLM in identifying both resource types while accommodating the variability in their implementations across different applications.

To handle large software systems with thousands of files and tens of thousands of functions and variables, we employ a COT-based staged filtering approach. As shown in Figure~\ref{fig:step2_prompt}, the LLM first examines metadata at the file level to make an initial decision about whether a file contains potential resources. If the LLM determines that more information is needed, it explicitly requests additional details about the classes, functions, and variables within the file to refine its analysis. This staged filtering approach reduces computational overhead while narrowing the search space systematically.

Figure~\ref{fig:step2_prompt} provides a snapshot of the prompts used in the class node stage of this process. The prompt includes a clear task summary and step-by-step instructions to guide the LLM in distinguishing between exclusive resources and shared resources. If the LLM cannot make a definitive decision based on the class name and description alone, it is prompted to explicitly request further information using the response format: “Answer ${INDEX}:NEED_DEF$ if you cannot decide simply based on the class name and description.”

Besides scalability, this CoT-based approach offers several advantages. First, it addresses \textit{comment loss} challenges in section~\ref{sec:design_opportunities} by delegating ambiguous decisions to subsequent stages. If a comment or description lacks sufficient detail, the LLM explicitly requests additional metadata, ensuring that missing information at one stage does not stop the identification of application resources. Second, the modularity of this design makes it highly portable and customizable. Each stage operates independently, allowing users to add or modify prompts for specific stages without impacting the overall workflow. For instance, users can introduce a new resource definition at the class node stage by appending their prompt to the existing instructions, leaving the rest of the process intact. 

\textbf{Step 3: finding operator functions for application resources.} 
After identifying potential application-defined resources, the next step is to locate operator functions that interact with them.  Operator functions provide crucial insights into the usage patterns of application resources and help identify potential points of contention. Our analysis reveals that resource usage primarily occurs in resource-related functions—such as member functions for resource objects or functions where resources are passed as parameters.

Based on this observation, we adopt a hybrid approach that combines static analysis and LLM-guided inference to systematically identify operator functions and resource usage patterns. In the static analysis phase, the data flow of each identified resource is analyzed to extract functions that potentially interact with the resource. The LLM then processes the function declaration and associated comments to evaluate whether the function represents a usage point. If the available information is insufficient, the LLM explicitly requests additional context, such as the function’s source code. Similar to how human developers analyze source code, the LLM inspects the function implementation to make a more informed decision about its role in resource usage.

Since different resource types exhibit distinct usage patterns, we provide the LLM with resource-specific operational definitions. For example, exclusive resources typically involve synchronization operations to prevent concurrent access. To classify such operations, we prompt the LLM with the concept of synchronization operations (e.g., LOCK, UNLOCK, and WAIT) and their typical code patterns. By tailoring prompts to the characteristics of each resource type, we enable the LLM to accurately classify resource operations and identify operator functions.

\textbf{Output.} By the end of this step, the LLM produces a list of potential resource usage points, categorized by their operations (e.g., synchronization for exclusive resources or concurrent access for shared resources). This information serves as input for the next static analysis module, which validates and refines the results to ensure accuracy.

\subsection{Static Analysis Validation}

The LLM module identifies resource candidates and their operator functions based on software metadata, providing broad coverage. However, this broad coverage often comes at the cost of low accuracy due to ambiguities and inconsistencies in application documentation. To address this, we developed a static analysis module that validates and refines the LLM’s output by applying stricter, code-based conditions to ensure the correctness of the identified resources and operator functions. This hybrid approach allows the static analysis to focus on verifying the LLM’s results with resource-specific validation patterns, ensuring both accuracy and explainability.

The static analysis module employs a dispatcher to classify each resource candidate by type and route it to a corresponding validation pass tailored to exclusive or shared resources.

\textit{Validation for exclusive resources.} For exclusive resources, contention occurs when a thread or process attempts to acquire a resource that is already held by another thread or process. Such contention is typically controlled through synchronization mechanisms, either system-level locks or custom application-level synchronization primitives. Both approaches require a thread to yield itself, either through a system call or assembly-level instructions.

The static analysis module begins with intra-procedural analysis to check whether the operator functions identified by the LLM contain synchronization primitives that cause the thread to yield. If a resource lacks such operator function involving synchronization, it is marked as invalid, as the absence of synchronization means the resource cannot block threads and, therefore, cannot cause resource contention.

Next, the module evaluates whether other usages of the exclusive resource correctly invoke the identified synchronization operator functions. It performs data flow analysis to identify the relationship between variables manipulated by the synchronization primitives and the resource-related variables identified by the LLM. If these variables do not share an assignment relationship, belong to the same data structure, or are explicitly marked as related by the LLM in step 2, the operator function is discarded.

\textit{Validation for shared resources.} For shared resources, contention often results from complex interactions with system-level resources, such as a buffer eviction when a write query attempts to store data in MySQL’s buffer pool. We observe that shared resources generally follow two common code patterns: (1) a control flow variable related to the resource creates divergent execution paths, where one path leads to fast execution and the other to slow execution, and (2) interactions with system-level resources, such as disk writes or queue operations, are integral to the resource’s operation.

To explore the first pattern, the static analysis module performs inter-procedural analysis to identify control flow variables in the operator function. It then checks whether these variables interact with the shared resource. If no interaction is found, the operator function is eliminated as a potential resource usage point.For second pattern, the module identifies system-level resource interactions by analyzing system calls within the operator functions. A whitelist of relevant system calls is defined for each type of shared resource. If the shared resource identified by the LLM does not interact with any system-level resource, the corresponding operator function is excluded as a valid usage point.

\textit{Scalability implications.}
Since the LLM module truncates the search space to a small set of application resource and their operator function. Our static analysis module can perform more comprehensive inter-procedural analysis without worrying about scalability. This allows the static analysis module to analyze the deep call stack to explore the resource usage pattern and validate the LLM module output.

%% file: section/design.tex
\section{Design of \tool}

We developed \tool, a practice profiler that enables \trace, to diagnose and explain application resource bottlenecks effectively. \tool leverages the hybrid resource identification module to integrate system-level and application-level resource tracing. To extend its capabilities, \tool introduces value-assisted data-flow profiling, enabling it to not only identify bottleneck resources but also explain why contention occurs from a code perspective. This combined approach provides developers with actionable insights into the root causes of performance issues.

\subsection{Overview of \tool}

\begin{figure}[t]
    \centering
      \includegraphics[width=2.8in]{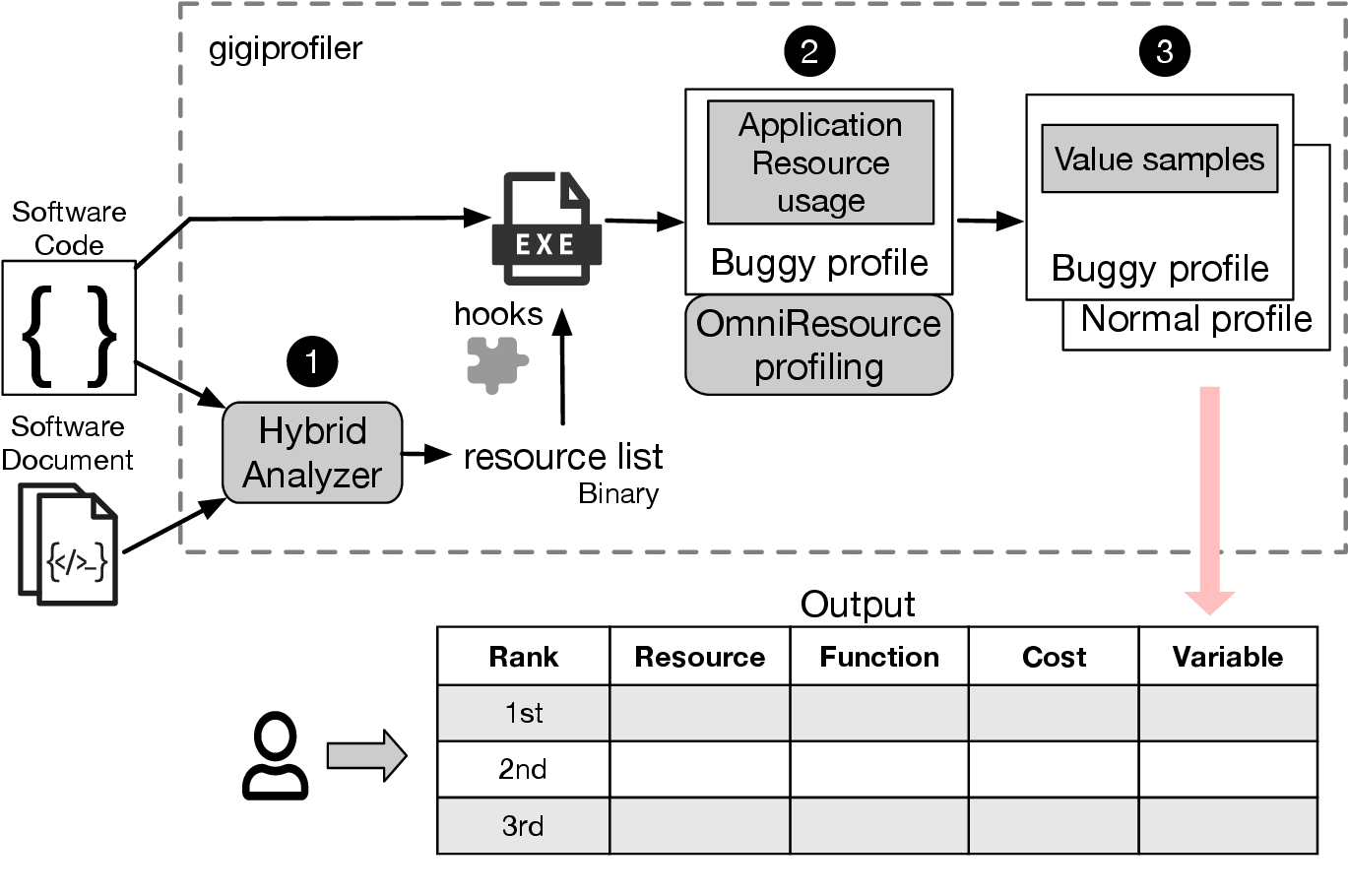}
      \caption{An overview of \tool.}
      \label{fig:overview_gigiprofiler}
  \end{figure}

Figure~\ref{fig:overview_gigiprofiler} shows the workflow of \tool, which operates in three main steps. First, software documentation and source code serve as input to the hybrid analyzer. The hybrid analyzer identifies potential application-defined resources and their associated operator functions, generating a comprehensive resource list. This list is then used to instrument the application binary by inserting hooks, enabling precise profiling of application resource usage. Second, using the instrumented binary, \tool profiles the application’s execution under buggy conditions. This profiling system monitors the usage patterns of application-defined resources and records the associated costs for each resource and operator function. This step generates a “buggy profile,” which identifies resource contention points, ranks resources by their impact on performance, and highlights candidate bottleneck resources. Third, to further investigate the root cause of resource contention, \tool performs a value-assisted data-flow analysis. This step compares variable values sampled during the buggy profile with those from a normal profile (executed under non-buggy conditions). By analyzing differences in these variable patterns, \tool provides insights into why resource contention occurs, pinpointing anomalous behaviors in the code. The final output of \tool is a detailed report that ranks resources, their associated functions, costs, and critical variables contributing to bottlenecks, offering actionable insights to resolve performance issues.

\subsection{Expose Application-Defined Resources Usage}

After identifying application-defined resources and their operator functions using the hybrid analyzer, the next step is to trace application resource usage. Tracing application-defined resources is latency-sensitive, as it operates within the critical path of software execution. To address this, \tool employs a lightweight instrumentation approach that minimizes the cost of tracing while maintaining accuracy.

The instrumentation is designed to record essential runtime metrics with minimal performance overhead, ensuring reliable profiling even under high workload scenarios.
For exclusive resources, \tool instruments the entry and terminator points of resource usage functions. These hooks record runtime metrics, such as resource usage time and execution time, during each invocation. For shared resources, \tool instruments synchronization primitives at the resource usage points. These hooks capture the times of resource acquisition and release, enabling the analysis of contention patterns.

To further reduce overhead, the \tool record the metrics directly to the thread’s local memory, avoiding costly interthread communication. The profiler periodically reads these metrics from each thread in the background to reconstruct the global view of resource usage, ensuring efficient data aggregation without disrupting application execution.

For tracing system-level resources and additional runtime data, such as call stacks, \tool integrates with \textit{perf} to periodically sample the information.

\input{section/algo.tex}

\subsection{Locating Performance Bottleneck}

Locating resource bottlenecks requires understanding how different resource types are used and determining which resources exhibit the highest contention. Since each resource type, exclusive or shared, has distinct usage patterns, \tool uses recorded runtime metrics to identify contention points and analyze the behavior of threads contributing to delays.

Algorithm~\ref{algo:locate} details \tool’s workflow for identifying resource bottlenecks. The process begins by iterating through all identified resources (\texttt{res\_list}) and their associated threads (\texttt{thread\_list}). For shared resources, \tool evaluates contention by checking if a thread is causing delays for others (\texttt{is\_contention}) or holding the resource excessively (\texttt{is\_hold}). For exclusive resources, it identifies blocking situations where threads fail to acquire a locked resource (\texttt{is\_blocked}) and calculates the time spent waiting or holding locks.

Once runtime data is collected, \tool identifies the bottleneck by determining the resource with the largest blocking time (\texttt{res\_block}). It further isolates the thread or process responsible for the contention (\texttt{longest\_holder}) by analyzing usage patterns. This systematic approach ensures that \tool can pinpoint the resource and thread contributing most to delays, enabling targeted performance optimization.

\subsection{Root Cause Analysis}

After identifying the bottleneck resource, the next critical question is why the resource contention occurs. This requires pinpointing the specific parts of the application code that misbehave under buggy conditions. For instance, if an operator function holds a resource for the same long duration in both buggy and normal executions, it is not the root cause. To locate the true root cause, \tool employs a comparison-based approach, analyzing differences between buggy and normal executions to identify abnormal behaviours. 

To minimize tracing overhead, \tool leverages static analysis to focus on key variables and code patterns that significantly impact performance, ensuring only the most relevant information is collected. \tool identifies loops within operator functions and their callers. These loops are critical as they often dictate how long a resource is held or how frequently it is accessed. During execution, \tool traces the variables associated with loop exit conditions and records the execution times of these loops under both buggy and normal conditions. By restricting tracing to these key variables and loops, \tool minimizes overhead while capturing the most impactful data. It then compares loop iteration counts and execution times between buggy and normal executions, flagging loops with significantly increased iteration counts or execution times under buggy conditions as potential root causes of resource contention.

For example, consider a database scenario where a background thread performs table backups, and a user thread periodically initiates update transactions. Under normal conditions, the background thread rarely competes for table resources. However, under specific isolation modes, buggy inputs may cause the background thread to repeatedly check the table’s status and perform backups, significantly increasing its resource holding time. \tool analyzes these loops and their callers to detect how buggy inputs trigger abnormal behavior, pinpointing the responsible variables.

%% file: section/algo.tex
\begin{algorithm}[t]
    \caption{Locate Resource Bottlenecks}
    \label{algo:locate}
    \SetKwInput{KwGlobal}{Global Vars}
    \KwGlobal{$res\_list$: List of resources}
    \KwGlobal{$thread\_list$: List of threads}
    \KwGlobal{$res\_usage$: Dict with {resource:usage} pairs}
    \KwGlobal{$res\_block$: Dict with {resource:block} pairs}
    \BlankLine
    
    \For{$res$ in $res\_list$}{
        \For{$thread$ in $thread\_list$}{
            \If{$res.type$ == \mbox{"shared"}}{
                \If{$is\_contention(thread, res)$}{
                    $contention\_time \gets get\_contention\_time(thread, res)$\;
                    $res\_usage[res] \gets res\_usage[res] + contention\_time$\;
                }
                \If{$is\_hold(thread, res)$}{
                    $hold\_amount \gets get\_hold\_amount(thread, res)$\;
                    $res\_block[res][thread] \gets hold\_amount$\;
                }
            }
            \If{$res.type$ == \mbox{"exclusive"}}{
                \If{$is\_blocked(thread, res)$}{
                    $blocked\_time \gets get\_blocked\_time(thread, res)$\;
                    $res\_block[res] \gets res\_block[res] + blocked\_time$\;
                }
                \If{$is\_hold(thread, res)$}{
                    $hold\_time \gets get\_hold\_time(thread, res)$\;
                    $res\_block[res][thread] \gets hold\_time$\;
                }
            }
        }
    }

    $bottleneck \gets \text{resource with the largest value in } res\_block$
    $max\_blocking\_time \gets res\_block[bottleneck]$\;
    $longest\_holder \gets \text{thread with the largest value in } res\_usage[bottleneck]$\;
\end{algorithm}

%% file: section/eval.tex
\section{Evaluation}

In evaluation, we address the following evaluation questions: (1) Can \tool effectively locate the root causes of performance problems? (2) Can \tool diagnose performance issues with previously unknown root causes? (3) What are the overhead of\tool? 

\input{section/case.tex}

\subsection{Experiment Setup}

All experiments were conducted on servers with 10-core Intel Xeon E5-2640 CPUs at 2.4 GHz, 64 GB DRAM, and a 480 GB SSD, running Ubuntu 20.04. We evaluated \tool on twelve real-world performance issues across five widely used applications: MySQL, PostgreSQL, MariaDB, Apache, and LLAMA. These applications are large, complex software with hundreds of thousands to millions of lines of code and exhibit performance issues that are challenging to diagnose.

The performance problems were collected from blog posts, official application forums, and bug trackers associated with the target applications. We selected issues using keywords like “performance,” “slow,” “degrade,” and resource-specific terms such as “buffer,” “memory,” and “log.” Issues were included if they were related to application resource contention and the reports provided sufficient information for reproduction. Bugs identified through straightforward code inspection were excluded, as they typically do not impact real users.

As shown in Table~\ref{table:description}, the twelve issues cover a range of symptoms and use cases, including table contention, inefficient caching, and thread or queue management problems. Diagnosing these cases is particularly challenging due to the complex interactions between application logic, custom resource management policies, and underlying hardware constraints. To evaluate \tool, we reproduced these issues based on the steps described in their reports. This process was nontrivial and time-consuming, as many issue reports were imprecisely described, requiring significant effort to replicate. Preparing the dataset took several person-months

For each issue, \tool was applied to the reproduction programs to diagnose the root causes. We compared \tool’s results with the root causes identified by developers in the issue reports. Additionally, we measured the runtime overhead incurred by \tool during the diagnosis process for each issue.

\subsection{Root Cause Identification}

\begin{table*}[!tbp]
    \setlength{\tabcolsep}{1.2ex}
    \footnotesize
    \centering
    \resizebox{\textwidth}{!}{
    \setcounter{magicrownumbers}{0}
    \begin{tabular}{@{}cccccc@{}}
    \toprule
    {\bf ID} & {\bf Resource Type} & {\bf Root Cause Function} & {\bf Root Cause Variable} & {\bf Analysis Time(s)} & {\bf Result in perf} \\
    \midrule
    c\rownumber & Exclusive resource   & row\_vers\_old\_has\_index\_entry (1st)     & undo\_rec    & 147.46s & 90th   \\
    c\rownumber & Exclusive resource       & wait\_for\_old\_version (1st)      & mdl\_context      &  86.4  & None    \\
    c\rownumber &  Exclusive resource  & lock\_table\_names  (1st), mysql\_inplace\_alter\_table(2nd)  & table\_version        & 102.53 s & None \\
    c\rownumber &Shared Resource   & buf\_page\_init\_for\_read(1st)      & buf\_pool\_t      & 89.70s  & None     \\
    c\rownumber & Shared Resource           & buf\_flush\_page(1st), buf\_LRU\_get\_free\_block(2nd)         & buf\_pool\_t       & 72.39  & 68(1st), 51(2nd)    \\
    \hline
    \hline
    c\rownumber & Exclusive resource           & dict\_acquire\_mdl\_shared(1st)         & mdl\_context       & 119.03 & None    \\
    c\rownumber & Shared Resource              & innobase\_srv\_conc\_enter\_innodb(1st)         & srv\_thread\_concurrency       & 67.52  & 14    \\
    \hline
    \hline
    c\rownumber & Exclusive resource             & GetMultiXactIdMembers(1st)         & LWLock       & 103.68  & 5    \\
    c\rownumber & Shared Resource             & pgss\_store(1st)         & LWLock       & 136.04  & 9    \\
    \hline
    \hline
    c\rownumber & Shared Resource            & start\_lingering\_close\_nonblocking(1st)         & event\_conn\_state\_t       & 35.20  & None    \\
    c\rownumber & Shared Resource              & ap\_queue\_info\_wait\_for\_idler(1st)         & fd\_queue\_info\_t       & 89.33  & None    \\
    \hline
    \hline
    c\rownumber & Exclusive resource            & process\_tasks(1st), next\_result(2nd)         & mutex       & 92.55 & None    \\
    \bottomrule
    \end{tabular}
    }
    \caption{Diagnosis results for the evaluated cases.}
    \vspace{-1em}
    \label{table:diagnosis_results}
\end{table*}

Table~\ref{table:diagnosis_results} summarizes the effectiveness of \tool in diagnosing 12 performance issues. The results demonstrate \tool’s robustness: it ranked the true root cause(s) as the top candidate in 9 out of 12 cases. For the remaining 3 cases, where multiple root causes contributed to the problem, \tool ranked all of them before other potential causes, ensuring comprehensive diagnostic accuracy.

Notably, all the issues lacked clear functional failures; they were reported due to degraded performance, such as increased execution time or higher memory consumption compared to baseline expectations. These cases are particularly challenging for tools designed to diagnose functional bugs but were effectively handled by \tool.

The analysis time was also efficient, especially when compared to the significant manual effort typically required for such diagnoses. On average, \tool required 95.15 seconds to identify the root cause, with no test exceeding 3 minutes of analysis time.
Below, we present detailed examples of how \tool diagnosed specific performance issues.

\textbf{Case 7: MDEV-24759.} This issue, affecting MariaDB versions prior to 10.5, comes from thread concurrency contention. MariaDB employs a queue to manage concurrent client requests. When the number of active clients exceeds the concurrency threshold, incoming requests are queued until previous ones are completed. This queuing mechanism creates a convoy effect, where a few long-running queries significantly degrade the overall server performance. \tool successfully identified this concurrency queue contention by detecting prolonged wait times for many queries in the \textit{innobase\_srv\_conc\_enter\_innodb} function, caused by the queue being dominated by long-running queries. \tool further revealed that the contention occurred when the \textit{srv\_thread\_concurrency} variable, which regulates the number of threads allowed to execute concurrently, was set lower than the number of active client requests.

\textbf{Case 10: Apache-60956.} This issue is caused by contention in the TCP send buffer. When a client prematurely stops receiving data during the asynchronous write completion process, the server continues writing to the send buffer, resulting in contention. The listening thread must wait for buffer space to become available before closing the connection after an idle timeout. \tool ranked buffer pool contention as the top candidate, identifying significantly prolonged execution in the \textit{start\_lingering\_close\_nonblocking} function due to waiting for free buffer space. Analysis of the \texttt{event\_conn\_state\_t} variable revealed that connections were in a waiting state for the buffer, differing from normal execution. Additionally, \tool traced the buffer consumption to buggy clients, pinpointing their role in exacerbating contention.

\subsection{Comparison with Profiling}

To evaluate \tool’s effectiveness, we compared its diagnostics against those generated by \textit{perf}, a widely used profiling tool. For each case, we used \textit{perf} to track the execution time of functions in blocked threads, ranking them in descending order. This comparison assesses whether the execution time provides meaningful insights for diagnosing resource contention in large-scale software.

As shown in table~\ref{table:diagnosis_results}, \tool outperformed \textit{perf}. In 7 cases, \textit{perf}  failed to record the root cause function, while \tool accurately identified the root cause. In 5 cases where \textit{perf} identified the root cause, the root cause function was absent from the top five functions ranked by \textit{perf}. 

\textit{Perf} was less effective because blocked threads often wait for resources, preventing \textit{perf} from capturing their execution time. For instance, in case 2, a long-running query held a metadata lock, creating an old version of the table. The backup transaction was blocked while waiting for the lock to be released. Since the backup thread yield itself during this period, \textit{perf} failed to capture its activity.

\subsection{Diagnosing Unresolved Performance Issues}

We further evaluated \tool’s capability to diagnose previously unresolved performance issues by applying it to two open cases in MariaDB. 

\textbf{MDEV-34989~\cite{mariadb_mdev_34989}.} This issue, introduced in version 11.7 with the vector search feature, caused significant performance regression. Developers observed that after using a SELECT statement to retrieve vectors from a table, subsequent INSERT operations into the same table would block indefinitely. Using \tool, we identified that the blocking occurred when the INSERT transaction attempted to synchronize updated data with the table while the table remained locked by other transactions. Specifically, \tool pinpointed the problematic code in \textit{vector\_mhnsw.cc} at line \textit{590}, where the insert thread waited indefinitely for the table to be released. Further analysis revealed that the SELECT transaction failed to release the table after acquiring it, as observed at line \textit{1296} in \textit{vector\_mhnsw.cc}. \tool traced the root cause to the \textit{mhnsw\_first} function, where the TABLE\_SHARE resource was not released when the table was empty. This premature return left subsequent transactions unable to acquire the TABLE\_SHARE resource, causing them to remain in a waiting state. Based on \tool’s findings, developers confirmed the diagnosis and implemented a fix to address the issue.

\textbf{MDEV-34836.~\cite{mariadb_mdev_34836}} This bug affects MariaDB version 10.6.0 and the root cause is Galera cluster’s handling of transactions. In a Galera cluster, when a row-level replication transaction(SR transaction) on a child table is executed, it locks both the child table and its parent table. If a table-level lock transaction is initiated on the parent table simultaneously, a conflict occurs. The DDL transaction on the parent table fails to properly abort the DML transaction on the child table due to Galera’s Total Order Isolation mechanism not handling foreign key dependency conflicts correctly. This failure leaves the DDL transaction on the parent table blocked until the DML transaction on the child table releases its locks. Using \tool, we identified the root cause and traced it back to Galera’s TOI mechanism. The diagnosis results were reported to MariaDB developers for further investigation.

\subsection{Overhead}

\tool has two components: an offline hybrid analyzer to identify application-defined resources and an online diagnostic system for tracing and diagnose resource bottleneck. We evaluated the overhead of both components in terms of offline preprocessing time and runtime diagnostic overhead.

For the offline hybrid analyzer, we applied it to MySQL, a large-scale application with millions of lines of code and approximately 9,200 files in its codebase. The scanning process was completed in 166 minutes: 130 minutes to identify application-defined resources and an additional 36 minutes to locate operator functions for each resource.

This efficiency is largely due to \tool’s organization of preprocessing data into a tree structure (Figure~\ref{fig:token_format}), which enables high parallelism during resource scanning. Each thread in \tool processes a subset of file nodes, enabling concurrent analysis of multiple files and significantly reducing the overall preprocessing time. Additionally, \tool’s CoT-based filtering prompt further optimizes the workflow by reducing the number of files requiring detailed inspection. Specifically, \tool analyzed only 4,621 out of 9,200 files to identify resources, and the search for operator functions reduced this number further to just 1,977 files.

\begin{figure}[t]
    \centering
    \includegraphics[width=3in]{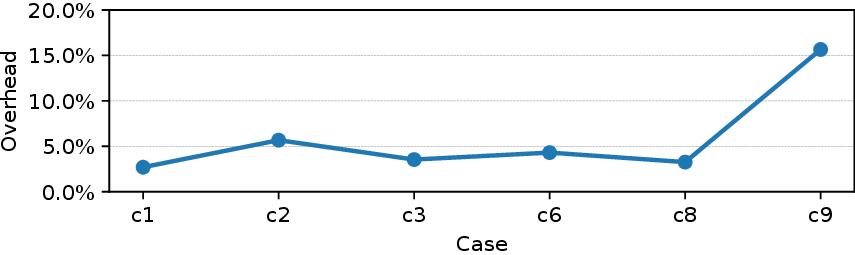}
    \caption{Overhead under 6 cases.}
    \label{fig:overhead}
\end{figure}

To measure runtime overhead, we compared the end-to-end throughput of software using \tool with baseline performance without \tool. We evaluated six cases, running each case five times and calculating the average overhead. Figure~\ref{fig:overhead} shows that \tool incurs an average runtime overhead of only 5.86\%, demonstrating its ability to diagnose performance issues with minimal impact on application performance.

\subsection{The Effectiveness of Hybrid Approach}

To evaluate the effectiveness of our hybrid approach, we compared it against using the LLM module and the static analysis module independently in identifying exclusive application resources in MySQL.

The static analysis module, relying on bottom-up control flow and data flow analysis, identified only 49 potential exclusive resources. Conversely, the LLM module identified over 2000 potential resources by leveraging high-level semantics extracted from comments and documentation. The hybrid approach refines the results to identify 92 resources.

To assess accuracy, we manually verified the resources identified by each approach. The LLM module exhibited a low accuracy of 3.7\%. The static analysis module achieved higher accuracy at 41.9\%. The hybrid approach, however, significantly improved accuracy to 80.4\%. This improvement comes from the hybrid approach’s design, where static analysis is focused on filtering false positives from the LLM’s output, allowing it to apply stricter validation rules without being limited by the low coverage of standalone static analysis.

%% file: section/case.tex
\begin{table*}[!tbp]
    \setlength{\tabcolsep}{1ex}
    \setcounter{magicrownumbers}{0}
    \footnotesize
    \centering
    \begin{tabular}{@{}lm{1.5cm}m{2cm}m{1.5cm}m{10cm}@{}}
     \toprule
      {\bf Id.} & {\bf App} & { \bf Resource} & {\bf Status} & {\bf Description} \\
     \midrule
      c\rownumber~(\href{https://bugs.mysql.com/bug.php?id=75540}{link}) & MySQL & UNDO log & Unresolved & Undo log occupies the buffer pool, blocking normal transactions during cleanup. \\
      c\rownumber~(\href{https://www.percona.com/blog/percona-xtrabackup-and-mysql-5-7-queries-in-waiting-for-table-flush-state/}{link}) & MySQL & Table metadata & Closed & A long-running query holds old version of table, forcing backup task to wait for its release. \\
      c\rownumber~(\href{https://bugs.mysql.com/bug.php?id=67647}{link}) & MySQL  & MDL lock & Closed & An uncommitted transaction blocks alter transactions that modify the table structure. \\
      c\rownumber~(\href{https://bugs.mysql.com/bug.php?id=96236}{link}) & MySQL  & Buffer pool  & Closed & Temporary tables fill the buffer pool, failing to allocate free pages for incoming transactions. \\
      c\rownumber~(\href{https://www.percona.com/blog/impact-of-swapping-on-mysql-performance/}{link}) & MySQL & Buffer pool & Closed & frequent buffer pool contention force new transactions to flush pages before proceeding. \\
    \hline
    \hline
      c\rownumber~(\href{https://jira.mariadb.org/browse/MDEV-34253}{link}) & MariaDB  & Foreign keys & Unresolved & Two alter transactions block each other due to conflicting foreign key constraints. \\
      c\rownumber~(\href{https://jira.mariadb.org/browse/MDEV-24759}{link}) & MariaDB & Queue & Closed & Excessive concurrent requests exceed the InnoDB thread limit, queuing in FIFO order. \\
    \hline
    \hline
      c\rownumber~(\href{https://www.postgresql.org/message-id/flat/2BEC2B3F-9B61-4C1D-9FB5-5FAB0F05EF86@yandex-team.ru}{link}) & PostgreSQL & MultiXact cache  & Closed & Cache flushes to disk frequently when MultiXact exceeds buffer capacity. \\
       c\rownumber~(\href{https://www.postgresql.org/message-id/flat/CAK-MWwSfx7SmUQ%3D_rB4q61sE8uP6LqH-1wM0DSYyBadjBe9BDw%40mail.gmail.com}{link}) & PostgreSQL & pgss &Closed & Low pg\_stat\_statements.max causes frequent contention and hash\_search failures. \\
    \hline
    \hline
      c\rownumber~(\href{https://bz.apache.org/bugzilla/show_bug.cgi?id=60956}{link}) & Apache  & Buffer & Closed & A full TCP buffer prevents the thread from closing SSL sessions, blocking other requests. \\
      c\rownumber~(\href{https://serverfault.com/questions/133561/apache-reaching-maxclients-and-locking-the-server}{link}) & Apache & Thread pool & Closed & The number of requests exceeds Apache's concurrency limit, causing new requests to wait. \\
    \hline
    \hline
      c\rownumber~(\href{https://github.com/ggerganov/llama.cpp/issues/4583}{link}) & LLAMA & Result queue & Closed & Tasks block each other in the result queue due to conflicting operations. \\
    \hline
    \bottomrule
    \end{tabular}
    \caption{Description of 12 \emph{real-world} performance issues we collected and reproduced in the five evaluated software systems.}
    \label{table:description}
    \vspace{-1em}
  \end{table*}

%% file: section/related.tex
\section{Related work}

Diagnosing performance bottlenecks in large-scale software systems has been a critical challenge in software engineering and systems research. Existing performance analysis tools can be broadly categorized into general profilers, causality analysis tools, and application-specific methods. 

\textbf{Performance Diagnosis.} Profilers~\cite{perf,gnu_gprof,systemtap,valgrind,Hangal2001perf,curtsinger2015coz,oprofile,gperftools} are fundamental tools for detecting performance problems and identifying associated symptoms. These tools focus on revealing performance-intensive events (e.g., functions, loops, or resource usage) but leave developers to manually infer the underlying reasons for resource contention.

Advanced profiling frameworks extend these capabilities by associating performance symptoms with their root causes. Critical path profiling\cite{Chow2014mystery,Szebenyi2009space} identifies the sequence of dependent tasks that dictate the overall execution time. Statistical debugging\cite{ren2019relational,libit2005scalable,Song2014statistical,Tanvir2021DMon,Burtsev2016abstractions} uses statistical analysis of program behavior to correlate performance anomalies with specific code regions or variables. While these tools excel in diagnosing system-level issues, they are not designed to handle application-defined resources, which often involve custom synchronization mechanisms and resource usage policies. 

There are also tools tailored for debugging specific types of performance problems by predefined patterns. For instance, loop-oriented debugging tools~\cite{Nistor2013Toddler,Song2017performance,Xiao2013delta} target performance degradation caused by inefficient or excessive loop iterations, while memory leak detectors~\cite{Vilk2018Bleak,Chang2022RESIN} focus on identifying and diagnosing memory usage issues. In contrast, \tool is designed to serve as a general-purpose diagnosis tool, capable of addressing application-defined resource contention.

\textbf{Causality Analysis.} Numerous advanced techniques have been developed for causality analysis in performance debugging~\cite{Jonathan2016Pivot,Attariyan2010ConfAid,Kasikci2015Failure,Kay2015Making,Lenin2012AppInsight,Yuan2010SherLog}. For example, Argus~\cite{Lingmei2021Argus} introduces causality graphs annotated with strong and weak edges, enabling prioritization of relational analyses for complex systems. Similarly Pensieve~\cite{Zhang2017Pensieve} demonstrate how runtime logs, combined with static analysis, can reconstruct causal paths and uncover performance bottlenecks. While these methods are effective in analyzing runtime interactions and identifying causal dependencies, they rely heavily on existing logs and annotations, which are often unavailable or incomplete for application-defined resources.

\textbf{Large Language Models for Software Analysis.} Modern large language models have revolutionized code-related tasks, including code generation~\cite{Du2024LLMClassEval,chen2021codex}, program repair~\cite{Yang2024CREF,Xia2023APR,sobania2023analysisautomaticbugfixing,Xia2024ChatRepair,Gao2020PatchGen}, automated testing and fuzzing~\cite{Kang2023LIBRO,Deng2024FuzzGPT}, and specification generation~\cite{ma2024specgenautomatedgenerationformal}. However, despite their advancements, LLMs face challenges in interpreting application-defined semantics due to their inherent variability. Recent works~\cite{su2024codeartbettercodemodels,tan2024llm4decompiledecompilingbinarycode,xie2024resym} combine LLMs with program analysis techniques, enabling applications like binary code understanding and decompilation. These efforts demonstrate LLMs’ potential for enhancing semantic comprehension in complex software

%% file: section/conclusion.tex
\section{Conclusion}

Diagnosing performance bottlenecks in large-scale software is challenging, particularly when addressing application-defined resource contention. We introduced \trace, an approach that integrates application-level resource tracing with system-level profiling. We realized this approach in \tool, which combines LLMs and static analysis to identify application-defined resources and operator functions. Through lightweight instrumentation and value-assisted profiling, \tool efficiently traces resource usage and identifies bottlenecks with minimal overhead and further locate the root causes by comparing buggy and normal execution data, identifying anomalous loops and key variables. Our evaluation demonstrates \tool’s effectiveness in diagnosing real-world performance.

%% file: paper.bbl

\begin{thebibliography}{60}


\ifx \showCODEN    \undefined \def \showCODEN     #1{\unskip}     \fi
\ifx \showDOI      \undefined \def \showDOI       #1{#1}\fi
\ifx \showISBNx    \undefined \def \showISBNx     #1{\unskip}     \fi
\ifx \showISBNxiii \undefined \def \showISBNxiii  #1{\unskip}     \fi
\ifx \showISSN     \undefined \def \showISSN      #1{\unskip}     \fi
\ifx \showLCCN     \undefined \def \showLCCN      #1{\unskip}     \fi
\ifx \shownote     \undefined \def \shownote      #1{#1}          \fi
\ifx \showarticletitle \undefined \def \showarticletitle #1{#1}   \fi
\ifx \showURL      \undefined \def \showURL       {\relax}        \fi
\providecommand\bibfield[2]{#2}
\providecommand\bibinfo[2]{#2}
\providecommand\natexlab[1]{#1}
\providecommand\showeprint[2][]{arXiv:#2}

\bibitem[Ahmed et~al\mbox{.}(2023)]%
        {Ahmed2023rootcause}
\bibfield{author}{\bibinfo{person}{Toufique Ahmed}, \bibinfo{person}{Supriyo
  Ghosh}, \bibinfo{person}{Chetan Bansal}, \bibinfo{person}{Thomas Zimmermann},
  \bibinfo{person}{Xuchao Zhang}, {and} \bibinfo{person}{Saravan Rajmohan}.}
  \bibinfo{year}{2023}\natexlab{}.
\newblock \showarticletitle{Recommending Root-Cause and Mitigation Steps for
  Cloud Incidents Using Large Language Models}. In
  \bibinfo{booktitle}{\emph{Proceedings of the 45th International Conference on
  Software Engineering}} (Melbourne, Victoria, Australia)
  \emph{(\bibinfo{series}{ICSE '23})}. \bibinfo{publisher}{IEEE Press},
  \bibinfo{pages}{1737–1749}.
\newblock
\showISBNx{9781665457019}
\urldef\tempurl%
\url{https://doi.org/10.1109/ICSE48619.2023.00149}
\showDOI{\tempurl}


\bibitem[Attariyan et~al\mbox{.}(2012)]%
        {attariyan2010xray}
\bibfield{author}{\bibinfo{person}{Mona Attariyan}, \bibinfo{person}{Michael
  Chow}, {and} \bibinfo{person}{Jason Flinn}.} \bibinfo{year}{2012}\natexlab{}.
\newblock \showarticletitle{X-ray: automating root-cause diagnosis of
  performance anomalies in production software}. In
  \bibinfo{booktitle}{\emph{Proceedings of the 10th USENIX Conference on
  Operating Systems Design and Implementation}} (Hollywood, CA, USA)
  \emph{(\bibinfo{series}{OSDI'12})}. \bibinfo{publisher}{USENIX Association},
  \bibinfo{address}{USA}, \bibinfo{pages}{307–320}.
\newblock
\showISBNx{9781931971966}


\bibitem[Attariyan and Flinn(2010)]%
        {Attariyan2010ConfAid}
\bibfield{author}{\bibinfo{person}{Mona Attariyan} {and} \bibinfo{person}{Jason
  Flinn}.} \bibinfo{year}{2010}\natexlab{}.
\newblock \showarticletitle{Automating configuration troubleshooting with
  dynamic information flow analysis}. In \bibinfo{booktitle}{\emph{Proceedings
  of the 9th USENIX Conference on Operating Systems Design and Implementation}}
  (Vancouver, BC, Canada) \emph{(\bibinfo{series}{OSDI'10})}.
  \bibinfo{publisher}{USENIX Association}, \bibinfo{address}{USA},
  \bibinfo{pages}{237–250}.
\newblock


\bibitem[Burtsev et~al\mbox{.}(2016)]%
        {Burtsev2016abstractions}
\bibfield{author}{\bibinfo{person}{Anton Burtsev}, \bibinfo{person}{David
  Johnson}, \bibinfo{person}{Mike Hibler}, \bibinfo{person}{Eric Eide}, {and}
  \bibinfo{person}{John Regehr}.} \bibinfo{year}{2016}\natexlab{}.
\newblock \showarticletitle{Abstractions for Practical Virtual Machine Replay}.
\newblock \bibinfo{journal}{\emph{SIGPLAN Not.}} \bibinfo{volume}{51},
  \bibinfo{number}{7} (\bibinfo{date}{March} \bibinfo{year}{2016}),
  \bibinfo{pages}{93–106}.
\newblock
\showISSN{0362-1340}
\urldef\tempurl%
\url{https://doi.org/10.1145/3007611.2892257}
\showDOI{\tempurl}


\bibitem[Chen et~al\mbox{.}(2021)]%
        {chen2021codex}
\bibfield{author}{\bibinfo{person}{Mark Chen}, \bibinfo{person}{Jerry Tworek},
  \bibinfo{person}{Heewoo Jun}, \bibinfo{person}{Qiming Yuan},
  \bibinfo{person}{Henrique~Ponde de Oliveira~Pinto}, \bibinfo{person}{Jared
  Kaplan}, \bibinfo{person}{Harri Edwards}, \bibinfo{person}{Yuri Burda},
  \bibinfo{person}{Nicholas Joseph}, \bibinfo{person}{Greg Brockman},
  \bibinfo{person}{Alex Ray}, \bibinfo{person}{Sandhini Puri},
  \bibinfo{person}{Gretchen Krueger}, \bibinfo{person}{Michael Petrov},
  \bibinfo{person}{Heidy Khlaaf}, \bibinfo{person}{Girish Sastry},
  \bibinfo{person}{Pamela Mishkin}, \bibinfo{person}{Brooke Chan},
  \bibinfo{person}{Scott Gray}, \bibinfo{person}{Nick Ryder},
  \bibinfo{person}{Michael Pavlov}, \bibinfo{person}{Alethea Power},
  \bibinfo{person}{Lukas Kaiser}, \bibinfo{person}{Mohammad Bavarian},
  \bibinfo{person}{Clemens Winter}, \bibinfo{person}{Philippe Tillet},
  \bibinfo{person}{Felipe~Petroski Such}, \bibinfo{person}{David Cummings},
  \bibinfo{person}{Matthias Plappert}, \bibinfo{person}{Fotios Chantzis},
  \bibinfo{person}{Elizabeth Barnes}, \bibinfo{person}{Ariel Herbert-Voss},
  \bibinfo{person}{William~H. Guss}, \bibinfo{person}{Alex Nichol},
  \bibinfo{person}{Andy Paino}, \bibinfo{person}{Nik Tezak},
  \bibinfo{person}{Jie Tang}, \bibinfo{person}{Igor Babuschkin},
  \bibinfo{person}{Suchir Balaji}, \bibinfo{person}{Shantanu Jain},
  \bibinfo{person}{William Saunders}, \bibinfo{person}{Christopher Hesse},
  \bibinfo{person}{Andreas Carr}, \bibinfo{person}{Jan Leike},
  \bibinfo{person}{Joshua Achiam}, \bibinfo{person}{Vedant Misra},
  \bibinfo{person}{Evan Morikawa}, \bibinfo{person}{Alec Radford},
  \bibinfo{person}{Matthew Knight}, \bibinfo{person}{Miles Brundage},
  \bibinfo{person}{Mira Murati}, \bibinfo{person}{Katja Mayer},
  \bibinfo{person}{Peter Welinder}, \bibinfo{person}{Bob McGrew},
  \bibinfo{person}{Dario Amodei}, \bibinfo{person}{Sam McCandlish},
  \bibinfo{person}{Ilya Sutskever}, {and} \bibinfo{person}{Wojciech Zaremba}.}
  \bibinfo{year}{2021}\natexlab{}.
\newblock \bibinfo{title}{Evaluating Large Language Models Trained on Code}.
\newblock
\newblock
\urldef\tempurl%
\url{https://openai.com/index/openai-codex/}
\showURL{%
\tempurl}


\bibitem[Chen et~al\mbox{.}(2024)]%
        {Chen2024rcacopilot}
\bibfield{author}{\bibinfo{person}{Yinfang Chen}, \bibinfo{person}{Huaibing
  Xie}, \bibinfo{person}{Minghua Ma}, \bibinfo{person}{Yu Kang},
  \bibinfo{person}{Xin Gao}, \bibinfo{person}{Liu Shi}, \bibinfo{person}{Yunjie
  Cao}, \bibinfo{person}{Xuedong Gao}, \bibinfo{person}{Hao Fan},
  \bibinfo{person}{Ming Wen}, \bibinfo{person}{Jun Zeng},
  \bibinfo{person}{Supriyo Ghosh}, \bibinfo{person}{Xuchao Zhang},
  \bibinfo{person}{Chaoyun Zhang}, \bibinfo{person}{Qingwei Lin},
  \bibinfo{person}{Saravan Rajmohan}, \bibinfo{person}{Dongmei Zhang}, {and}
  \bibinfo{person}{Tianyin Xu}.} \bibinfo{year}{2024}\natexlab{}.
\newblock \showarticletitle{Automatic Root Cause Analysis via Large Language
  Models for Cloud Incidents}. In \bibinfo{booktitle}{\emph{Proceedings of the
  Nineteenth European Conference on Computer Systems}} (Athens, Greece)
  \emph{(\bibinfo{series}{EuroSys '24})}. \bibinfo{publisher}{Association for
  Computing Machinery}, \bibinfo{address}{New York, NY, USA},
  \bibinfo{pages}{674–688}.
\newblock
\showISBNx{9798400704376}
\urldef\tempurl%
\url{https://doi.org/10.1145/3627703.3629553}
\showDOI{\tempurl}


\bibitem[Chow et~al\mbox{.}(2014)]%
        {Chow2014mystery}
\bibfield{author}{\bibinfo{person}{Michael Chow}, \bibinfo{person}{David
  Meisner}, \bibinfo{person}{Jason Flinn}, \bibinfo{person}{Daniel Peek}, {and}
  \bibinfo{person}{Thomas~F. Wenisch}.} \bibinfo{year}{2014}\natexlab{}.
\newblock \showarticletitle{The mystery machine: end-to-end performance
  analysis of large-scale internet services}. In
  \bibinfo{booktitle}{\emph{Proceedings of the 11th USENIX Conference on
  Operating Systems Design and Implementation}} (Broomfield, CO)
  \emph{(\bibinfo{series}{OSDI'14})}. \bibinfo{publisher}{USENIX Association},
  \bibinfo{address}{USA}, \bibinfo{pages}{217–231}.
\newblock
\showISBNx{9781931971164}


\bibitem[Cloud(2019)]%
        {alibaba_undo_logs}
\bibfield{author}{\bibinfo{person}{Alibaba Cloud}.}
  \bibinfo{year}{2019}\natexlab{}.
\newblock \bibinfo{title}{An In-Depth Analysis of Undo Logs in InnoDB}.
\newblock
\newblock
\urldef\tempurl%
\url{https://www.alibabacloud.com/blog/an-in-depth-analysis-of-undo-logs-in-innodb_598966}
\showURL{%
\tempurl}


\bibitem[Curtsinger and Berger(2015)]%
        {curtsinger2015coz}
\bibfield{author}{\bibinfo{person}{Charlie Curtsinger} {and}
  \bibinfo{person}{Emery~D. Berger}.} \bibinfo{year}{2015}\natexlab{}.
\newblock \showarticletitle{Coz: finding code that counts with causal
  profiling}. In \bibinfo{booktitle}{\emph{Proceedings of the 25th Symposium on
  Operating Systems Principles}} (Monterey, California)
  \emph{(\bibinfo{series}{SOSP '15})}. \bibinfo{publisher}{Association for
  Computing Machinery}, \bibinfo{address}{New York, NY, USA},
  \bibinfo{pages}{184–197}.
\newblock
\showISBNx{9781450338349}
\urldef\tempurl%
\url{https://doi.org/10.1145/2815400.2815409}
\showDOI{\tempurl}


\bibitem[Deng et~al\mbox{.}(2024)]%
        {Deng2024FuzzGPT}
\bibfield{author}{\bibinfo{person}{Yinlin Deng},
  \bibinfo{person}{Chunqiu~Steven Xia}, \bibinfo{person}{Chenyuan Yang},
  \bibinfo{person}{Shizhuo~Dylan Zhang}, \bibinfo{person}{Shujing Yang}, {and}
  \bibinfo{person}{Lingming Zhang}.} \bibinfo{year}{2024}\natexlab{}.
\newblock \showarticletitle{Large Language Models are Edge-Case Generators:
  Crafting Unusual Programs for Fuzzing Deep Learning Libraries}. In
  \bibinfo{booktitle}{\emph{Proceedings of the IEEE/ACM 46th International
  Conference on Software Engineering}} (Lisbon, Portugal)
  \emph{(\bibinfo{series}{ICSE '24})}. \bibinfo{publisher}{Association for
  Computing Machinery}, \bibinfo{address}{New York, NY, USA}, Article
  \bibinfo{articleno}{70}, \bibinfo{numpages}{13}~pages.
\newblock
\showISBNx{9798400702174}
\urldef\tempurl%
\url{https://doi.org/10.1145/3597503.3623343}
\showDOI{\tempurl}


\bibitem[Developers(2023a)]%
        {oprofile}
\bibfield{author}{\bibinfo{person}{OProfile Developers}.}
  \bibinfo{year}{2023}\natexlab{a}.
\newblock \bibinfo{title}{OProfile: A System Profiler for Linux}.
\newblock
\newblock
\urldef\tempurl%
\url{https://oprofile.sourceforge.io/about/}
\showURL{%
\tempurl}


\bibitem[Developers(2023b)]%
        {systemtap}
\bibfield{author}{\bibinfo{person}{SystemTap Developers}.}
  \bibinfo{year}{2023}\natexlab{b}.
\newblock \bibinfo{title}{SystemTap: Simplifying System-Level Observability}.
\newblock
\newblock
\urldef\tempurl%
\url{https://sourceware.org/systemtap/}
\showURL{%
\tempurl}


\bibitem[Developers(2023c)]%
        {valgrind}
\bibfield{author}{\bibinfo{person}{Valgrind Developers}.}
  \bibinfo{year}{2023}\natexlab{c}.
\newblock \bibinfo{title}{Valgrind: Instrumentation Framework for Building
  Dynamic Analysis Tools}.
\newblock
\newblock
\urldef\tempurl%
\url{https://valgrind.org/}
\showURL{%
\tempurl}


\bibitem[Du et~al\mbox{.}(2024)]%
        {Du2024LLMClassEval}
\bibfield{author}{\bibinfo{person}{Xueying Du}, \bibinfo{person}{Mingwei Liu},
  \bibinfo{person}{Kaixin Wang}, \bibinfo{person}{Hanlin Wang},
  \bibinfo{person}{Junwei Liu}, \bibinfo{person}{Yixuan Chen},
  \bibinfo{person}{Jiayi Feng}, \bibinfo{person}{Chaofeng Sha},
  \bibinfo{person}{Xin Peng}, {and} \bibinfo{person}{Yiling Lou}.}
  \bibinfo{year}{2024}\natexlab{}.
\newblock \showarticletitle{Evaluating Large Language Models in Class-Level
  Code Generation}. In \bibinfo{booktitle}{\emph{Proceedings of the IEEE/ACM
  46th International Conference on Software Engineering}} (Lisbon, Portugal)
  \emph{(\bibinfo{series}{ICSE '24})}. \bibinfo{publisher}{Association for
  Computing Machinery}, \bibinfo{address}{New York, NY, USA}, Article
  \bibinfo{articleno}{81}, \bibinfo{numpages}{13}~pages.
\newblock
\showISBNx{9798400702174}
\urldef\tempurl%
\url{https://doi.org/10.1145/3597503.3639219}
\showDOI{\tempurl}


\bibitem[Foundation({[n.\,d.]})]%
        {perf}
\bibfield{author}{\bibinfo{person}{Linux Foundation}.}
  \bibinfo{year}{[n.\,d.]}\natexlab{}.
\newblock \bibinfo{title}{Perf: Linux profiling with performance counters}.
\newblock
\newblock
\urldef\tempurl%
\url{https://perf.wiki.kernel.org/}
\showURL{%
\tempurl}


\bibitem[Gao and Roychoudhury(2020)]%
        {Gao2020PatchGen}
\bibfield{author}{\bibinfo{person}{Xiang Gao} {and} \bibinfo{person}{Abhik
  Roychoudhury}.} \bibinfo{year}{2020}\natexlab{}.
\newblock \showarticletitle{Interactive Patch Generation and Suggestion}. In
  \bibinfo{booktitle}{\emph{Proceedings of the IEEE/ACM 42nd International
  Conference on Software Engineering Workshops}} (Seoul, Republic of Korea)
  \emph{(\bibinfo{series}{ICSEW'20})}. \bibinfo{publisher}{Association for
  Computing Machinery}, \bibinfo{address}{New York, NY, USA},
  \bibinfo{pages}{17–18}.
\newblock
\showISBNx{9781450379632}
\urldef\tempurl%
\url{https://doi.org/10.1145/3387940.3392179}
\showDOI{\tempurl}


\bibitem[gperftools Developers(2023)]%
        {gperftools}
\bibfield{author}{\bibinfo{person}{gperftools Developers}.}
  \bibinfo{year}{2023}\natexlab{}.
\newblock \bibinfo{title}{gperftools: Google Performance Tools}.
\newblock
\newblock
\urldef\tempurl%
\url{https://github.com/gperftools/gperftools}
\showURL{%
\tempurl}


\bibitem[Hangal and Lam(2002)]%
        {Hangal2001perf}
\bibfield{author}{\bibinfo{person}{Sudheendra Hangal} {and}
  \bibinfo{person}{Monica~S. Lam}.} \bibinfo{year}{2002}\natexlab{}.
\newblock \showarticletitle{Tracking down software bugs using automatic anomaly
  detection}. In \bibinfo{booktitle}{\emph{Proceedings of the 24th
  International Conference on Software Engineering}} (Orlando, Florida)
  \emph{(\bibinfo{series}{ICSE '02})}. \bibinfo{publisher}{Association for
  Computing Machinery}, \bibinfo{address}{New York, NY, USA},
  \bibinfo{pages}{291–301}.
\newblock
\showISBNx{158113472X}
\urldef\tempurl%
\url{https://doi.org/10.1145/581339.581377}
\showDOI{\tempurl}


\bibitem[Hu et~al\mbox{.}(2023)]%
        {hu23pbox}
\bibfield{author}{\bibinfo{person}{Yigong Hu}, \bibinfo{person}{Gongqi Huang},
  {and} \bibinfo{person}{Peng Huang}.} \bibinfo{year}{2023}\natexlab{}.
\newblock \showarticletitle{Pushing Performance Isolation Boundaries into
  Application with pBox}. In \bibinfo{booktitle}{\emph{Proceedings of the 29th
  Symposium on Operating Systems Principles}} (Koblenz, Germany)
  \emph{(\bibinfo{series}{SOSP '23})}. \bibinfo{publisher}{Association for
  Computing Machinery}, \bibinfo{address}{New York, NY, USA},
  \bibinfo{pages}{247--263}.
\newblock
\showISBNx{9798400702297}
\urldef\tempurl%
\url{https://doi.org/10.1145/3600006.3613159}
\showDOI{\tempurl}


\bibitem[Kang et~al\mbox{.}(2023)]%
        {Kang2023LIBRO}
\bibfield{author}{\bibinfo{person}{Sungmin Kang}, \bibinfo{person}{Juyeon
  Yoon}, {and} \bibinfo{person}{Shin Yoo}.} \bibinfo{year}{2023}\natexlab{}.
\newblock \showarticletitle{Large Language Models are Few-Shot Testers:
  Exploring LLM-Based General Bug Reproduction}. In
  \bibinfo{booktitle}{\emph{Proceedings of the 45th International Conference on
  Software Engineering}} (Melbourne, Victoria, Australia)
  \emph{(\bibinfo{series}{ICSE '23})}. \bibinfo{publisher}{IEEE Press},
  \bibinfo{pages}{2312–2323}.
\newblock
\showISBNx{9781665457019}
\urldef\tempurl%
\url{https://doi.org/10.1109/ICSE48619.2023.00194}
\showDOI{\tempurl}


\bibitem[Kasikci et~al\mbox{.}(2015)]%
        {Kasikci2015Failure}
\bibfield{author}{\bibinfo{person}{Baris Kasikci}, \bibinfo{person}{Benjamin
  Schubert}, \bibinfo{person}{Cristiano Pereira}, \bibinfo{person}{Gilles
  Pokam}, {and} \bibinfo{person}{George Candea}.}
  \bibinfo{year}{2015}\natexlab{}.
\newblock \showarticletitle{Failure sketching: a technique for automated root
  cause diagnosis of in-production failures}. In
  \bibinfo{booktitle}{\emph{Proceedings of the 25th Symposium on Operating
  Systems Principles}} (Monterey, California) \emph{(\bibinfo{series}{SOSP
  '15})}. \bibinfo{publisher}{Association for Computing Machinery},
  \bibinfo{address}{New York, NY, USA}, \bibinfo{pages}{344–360}.
\newblock
\showISBNx{9781450338349}
\urldef\tempurl%
\url{https://doi.org/10.1145/2815400.2815412}
\showDOI{\tempurl}


\bibitem[Khan et~al\mbox{.}(2021)]%
        {Tanvir2021DMon}
\bibfield{author}{\bibinfo{person}{Tanvir~Ahmed Khan}, \bibinfo{person}{Ian
  Neal}, \bibinfo{person}{Gilles Pokam}, \bibinfo{person}{Barzan Mozafari},
  {and} \bibinfo{person}{Baris Kasikci}.} \bibinfo{year}{2021}\natexlab{}.
\newblock \showarticletitle{DMon: Efficient Detection and Correction of Data
  Locality Problems Using Selective Profiling}. In
  \bibinfo{booktitle}{\emph{15th {USENIX} Symposium on Operating Systems Design
  and Implementation ({OSDI} 21)}}. \bibinfo{publisher}{{USENIX} Association},
  \bibinfo{pages}{163--181}.
\newblock
\showISBNx{978-1-939133-22-9}
\urldef\tempurl%
\url{https://www.usenix.org/conference/osdi21/presentation/khan}
\showURL{%
\tempurl}


\bibitem[Liblit et~al\mbox{.}(2005)]%
        {libit2005scalable}
\bibfield{author}{\bibinfo{person}{Ben Liblit}, \bibinfo{person}{Mayur Naik},
  \bibinfo{person}{Alice~X. Zheng}, \bibinfo{person}{Alex Aiken}, {and}
  \bibinfo{person}{Michael~I. Jordan}.} \bibinfo{year}{2005}\natexlab{}.
\newblock \showarticletitle{Scalable statistical bug isolation}. In
  \bibinfo{booktitle}{\emph{Proceedings of the 2005 ACM SIGPLAN Conference on
  Programming Language Design and Implementation}} (Chicago, IL, USA)
  \emph{(\bibinfo{series}{PLDI '05})}. \bibinfo{publisher}{Association for
  Computing Machinery}, \bibinfo{address}{New York, NY, USA},
  \bibinfo{pages}{15–26}.
\newblock
\showISBNx{1595930566}
\urldef\tempurl%
\url{https://doi.org/10.1145/1065010.1065014}
\showDOI{\tempurl}


\bibitem[Lou et~al\mbox{.}(2022)]%
        {Chang2022RESIN}
\bibfield{author}{\bibinfo{person}{Chang Lou}, \bibinfo{person}{Cong Chen},
  \bibinfo{person}{Peng Huang}, \bibinfo{person}{Yingnong Dang},
  \bibinfo{person}{Si Qin}, \bibinfo{person}{Xinsheng Yang},
  \bibinfo{person}{Xukun Li}, \bibinfo{person}{Qingwei Lin}, {and}
  \bibinfo{person}{Murali Chintalapati}.} \bibinfo{year}{2022}\natexlab{}.
\newblock \showarticletitle{{RESIN}: A Holistic Service for Dealing with Memory
  Leaks in Production Cloud Infrastructure}. In \bibinfo{booktitle}{\emph{16th
  USENIX Symposium on Operating Systems Design and Implementation (OSDI 22)}}.
  \bibinfo{publisher}{USENIX Association}, \bibinfo{address}{Carlsbad, CA},
  \bibinfo{pages}{109--125}.
\newblock
\showISBNx{978-1-939133-28-1}
\urldef\tempurl%
\url{https://www.usenix.org/conference/osdi22/presentation/lou-resin}
\showURL{%
\tempurl}


\bibitem[Ma et~al\mbox{.}(2024)]%
        {ma2024specgenautomatedgenerationformal}
\bibfield{author}{\bibinfo{person}{Lezhi Ma}, \bibinfo{person}{Shangqing Liu},
  \bibinfo{person}{Yi Li}, \bibinfo{person}{Xiaofei Xie}, {and}
  \bibinfo{person}{Lei Bu}.} \bibinfo{year}{2024}\natexlab{}.
\newblock \bibinfo{title}{SpecGen: Automated Generation of Formal Program
  Specifications via Large Language Models}.
\newblock
\newblock
\showeprint[arxiv]{2401.08807}~[cs.SE]
\urldef\tempurl%
\url{https://arxiv.org/abs/2401.08807}
\showURL{%
\tempurl}


\bibitem[Mace et~al\mbox{.}(2016)]%
        {Jonathan2016Pivot}
\bibfield{author}{\bibinfo{person}{Jonathan Mace}, \bibinfo{person}{Ryan
  Roelke}, {and} \bibinfo{person}{Rodrigo Fonseca}.}
  \bibinfo{year}{2016}\natexlab{}.
\newblock \showarticletitle{Pivot Tracing: Dynamic Causal Monitoring for
  Distributed Systems}. In \bibinfo{booktitle}{\emph{2016 USENIX Annual
  Technical Conference (USENIX ATC 16)}}. \bibinfo{publisher}{USENIX
  Association}, \bibinfo{address}{Denver, CO}.
\newblock
\urldef\tempurl%
\url{https://www.usenix.org/conference/atc16/technical-sessions/presentation/mace}
\showURL{%
\tempurl}


\bibitem[MariaDB(2023a)]%
        {mariadb_mdev_34836}
\bibfield{author}{\bibinfo{person}{MariaDB}.} \bibinfo{year}{2023}\natexlab{a}.
\newblock \bibinfo{title}{MDEV-34836: Performance regression due to excessive
  purge lag}.
\newblock
\newblock
\urldef\tempurl%
\url{https://jira.mariadb.org/browse/MDEV-34836}
\showURL{%
\tempurl}
\newblock
\shownote{Accessed: 2023-10-01}.


\bibitem[MariaDB(2023b)]%
        {mariadb_mdev_34989}
\bibfield{author}{\bibinfo{person}{MariaDB}.} \bibinfo{year}{2023}\natexlab{b}.
\newblock \bibinfo{title}{MDEV-34989: Performance regression due to excessive
  purge lag}.
\newblock
\newblock
\urldef\tempurl%
\url{https://jira.mariadb.org/browse/MDEV-34989}
\showURL{%
\tempurl}
\newblock
\shownote{Accessed: 2023-10-01}.


\bibitem[{MySQL}(2015)]%
        {mysql_bug_75540}
\bibfield{author}{\bibinfo{person}{{MySQL}}.} \bibinfo{year}{2015}\natexlab{}.
\newblock \bibinfo{title}{MySQL Bug \#75540: Purge thread causes performance
  regression}.
\newblock
\newblock
\urldef\tempurl%
\url{https://bugs.mysql.com/bug.php?id=75540}
\showURL{%
\tempurl}


\bibitem[{MySQL}(2020)]%
        {mysql_bug_99315}
\bibfield{author}{\bibinfo{person}{{MySQL}}.} \bibinfo{year}{2020}\natexlab{}.
\newblock \bibinfo{title}{MySQL Bug \#99315: Performance regression due to
  excessive purge lag}.
\newblock
\newblock
\urldef\tempurl%
\url{https://bugs.mysql.com/bug.php?id=99315}
\showURL{%
\tempurl}


\bibitem[Nistor et~al\mbox{.}(2013)]%
        {Nistor2013Toddler}
\bibfield{author}{\bibinfo{person}{Adrian Nistor}, \bibinfo{person}{Linhai
  Song}, \bibinfo{person}{Darko Marinov}, {and} \bibinfo{person}{Shan Lu}.}
  \bibinfo{year}{2013}\natexlab{}.
\newblock \showarticletitle{Toddler: detecting performance problems via similar
  memory-access patterns}. In \bibinfo{booktitle}{\emph{Proceedings of the 2013
  International Conference on Software Engineering}} (San Francisco, CA, USA)
  \emph{(\bibinfo{series}{ICSE '13})}. \bibinfo{publisher}{IEEE Press},
  \bibinfo{pages}{562–571}.
\newblock
\showISBNx{9781467330763}


\bibitem[Ousterhout et~al\mbox{.}(2015)]%
        {Kay2015Making}
\bibfield{author}{\bibinfo{person}{Kay Ousterhout}, \bibinfo{person}{Ryan
  Rasti}, \bibinfo{person}{Sylvia Ratnasamy}, \bibinfo{person}{Scott Shenker},
  {and} \bibinfo{person}{Byung-Gon Chun}.} \bibinfo{year}{2015}\natexlab{}.
\newblock \showarticletitle{Making Sense of Performance in Data Analytics
  Frameworks}. In \bibinfo{booktitle}{\emph{12th USENIX Symposium on Networked
  Systems Design and Implementation (NSDI 15)}}. \bibinfo{publisher}{USENIX
  Association}, \bibinfo{address}{Oakland, CA}, \bibinfo{pages}{293--307}.
\newblock
\showISBNx{978-1-931971-218}
\urldef\tempurl%
\url{https://www.usenix.org/conference/nsdi15/technical-sessions/presentation/ousterhout}
\showURL{%
\tempurl}


\bibitem[Percona(2014a)]%
        {percona_transaction_history}
\bibfield{author}{\bibinfo{person}{Percona}.} \bibinfo{year}{2014}\natexlab{a}.
\newblock \bibinfo{title}{InnoDB Transaction History Often Hides Dangerous
  Debt}.
\newblock
\newblock
\urldef\tempurl%
\url{https://www.percona.com/blog/innodb-transaction-history-often-hides-dangerous-debt/?utm_content=buffer42e4a&utm_medium=social&utm_source=app.net&utm_campaign=buffer}
\showURL{%
\tempurl}


\bibitem[Percona(2014b)]%
        {percona_multiversioning}
\bibfield{author}{\bibinfo{person}{Percona}.} \bibinfo{year}{2014}\natexlab{b}.
\newblock \bibinfo{title}{InnoDB's Multi-Versioning Handling Can Be Achilles'
  Heel}.
\newblock
\newblock
\urldef\tempurl%
\url{https://www.percona.com/blog/innodbs-multi-versioning-handling-can-be-achilles-heel/}
\showURL{%
\tempurl}


\bibitem[Percona(2015)]%
        {percona_isolation_modes}
\bibfield{author}{\bibinfo{person}{Percona}.} \bibinfo{year}{2015}\natexlab{}.
\newblock \bibinfo{title}{MySQL Performance Implications of InnoDB Isolation
  Modes}.
\newblock
\newblock
\urldef\tempurl%
\url{https://www.percona.com/blog/mysql-performance-implications-of-innodb-isolation-modes/}
\showURL{%
\tempurl}


\bibitem[Percona(2017a)]%
        {percona_hung_transaction}
\bibfield{author}{\bibinfo{person}{Percona}.} \bibinfo{year}{2017}\natexlab{a}.
\newblock \bibinfo{title}{Chasing a Hung Transaction in MySQL: InnoDB History
  Length Strikes Back}.
\newblock
\newblock
\urldef\tempurl%
\url{https://www.percona.com/blog/chasing-a-hung-transaction-in-mysql-innodb-history-length-strikes-back/}
\showURL{%
\tempurl}


\bibitem[Percona(2017b)]%
        {percona_swapping}
\bibfield{author}{\bibinfo{person}{Percona}.} \bibinfo{year}{2017}\natexlab{b}.
\newblock \bibinfo{title}{Impact of Swapping on MySQL Performance}.
\newblock
\newblock
\urldef\tempurl%
\url{https://www.percona.com/blog/impact-of-swapping-on-mysql-performance/}
\showURL{%
\tempurl}


\bibitem[Percona(2023)]%
        {percona_xtrabackup}
\bibfield{author}{\bibinfo{person}{Percona}.} \bibinfo{year}{2023}\natexlab{}.
\newblock \bibinfo{title}{Percona XtraBackup and MySQL 5.7: Queries in "Waiting
  for table flush" State}.
\newblock
\newblock
\urldef\tempurl%
\url{https://www.percona.com/blog/percona-xtrabackup-and-mysql-5-7-queries-in-waiting-for-table-flush-state/}
\showURL{%
\tempurl}


\bibitem[Project(1998)]%
        {gnu_gprof}
\bibfield{author}{\bibinfo{person}{GNU Project}.}
  \bibinfo{year}{1998}\natexlab{}.
\newblock \bibinfo{booktitle}{\emph{GNU gprof: a Call Graph Execution
  Profiler}}.
\newblock
\urldef\tempurl%
\url{https://ftp.gnu.org/old-gnu/Manuals/gprof-2.9.1/html_mono/gprof.html}
\showURL{%
\tempurl}


\bibitem[Ravindranath et~al\mbox{.}(2012)]%
        {Lenin2012AppInsight}
\bibfield{author}{\bibinfo{person}{Lenin Ravindranath},
  \bibinfo{person}{Jitendra Padhye}, \bibinfo{person}{Sharad Agarwal},
  \bibinfo{person}{Ratul Mahajan}, \bibinfo{person}{Ian Obermiller}, {and}
  \bibinfo{person}{Shahin Shayandeh}.} \bibinfo{year}{2012}\natexlab{}.
\newblock \showarticletitle{{AppInsight}: Mobile App Performance Monitoring in
  the Wild}. In \bibinfo{booktitle}{\emph{10th USENIX Symposium on Operating
  Systems Design and Implementation (OSDI 12)}}. \bibinfo{publisher}{USENIX
  Association}, \bibinfo{address}{Hollywood, CA}, \bibinfo{pages}{107--120}.
\newblock
\showISBNx{978-1-931971-96-6}
\urldef\tempurl%
\url{https://www.usenix.org/conference/osdi12/technical-sessions/presentation/ravindranath}
\showURL{%
\tempurl}


\bibitem[Ren et~al\mbox{.}(2023)]%
        {ren2019relational}
\bibfield{author}{\bibinfo{person}{Xiang~(Jenny) Ren}, \bibinfo{person}{Sitao
  Wang}, \bibinfo{person}{Zhuqi Jin}, \bibinfo{person}{David Lion},
  \bibinfo{person}{Adrian Chiu}, \bibinfo{person}{Tianyin Xu}, {and}
  \bibinfo{person}{Ding Yuan}.} \bibinfo{year}{2023}\natexlab{}.
\newblock \showarticletitle{Relational Debugging --- Pinpointing Root Causes of
  Performance Problems}. In \bibinfo{booktitle}{\emph{17th USENIX Symposium on
  Operating Systems Design and Implementation (OSDI 23)}}.
  \bibinfo{publisher}{USENIX Association}, \bibinfo{address}{Boston, MA},
  \bibinfo{pages}{65--80}.
\newblock
\showISBNx{978-1-939133-34-2}
\urldef\tempurl%
\url{https://www.usenix.org/conference/osdi23/presentation/ren}
\showURL{%
\tempurl}


\bibitem[Sobania et~al\mbox{.}(2023)]%
        {sobania2023analysisautomaticbugfixing}
\bibfield{author}{\bibinfo{person}{Dominik Sobania}, \bibinfo{person}{Martin
  Briesch}, \bibinfo{person}{Carol Hanna}, {and} \bibinfo{person}{Justyna
  Petke}.} \bibinfo{year}{2023}\natexlab{}.
\newblock \bibinfo{title}{An Analysis of the Automatic Bug Fixing Performance
  of ChatGPT}.
\newblock
\newblock
\showeprint[arxiv]{2301.08653}~[cs.SE]
\urldef\tempurl%
\url{https://arxiv.org/abs/2301.08653}
\showURL{%
\tempurl}


\bibitem[Song and Lu(2014)]%
        {Song2014statistical}
\bibfield{author}{\bibinfo{person}{Linhai Song} {and} \bibinfo{person}{Shan
  Lu}.} \bibinfo{year}{2014}\natexlab{}.
\newblock \showarticletitle{Statistical debugging for real-world performance
  problems}.
\newblock \bibinfo{journal}{\emph{SIGPLAN Not.}} \bibinfo{volume}{49},
  \bibinfo{number}{10} (\bibinfo{date}{Oct.} \bibinfo{year}{2014}),
  \bibinfo{pages}{561–578}.
\newblock
\showISSN{0362-1340}
\urldef\tempurl%
\url{https://doi.org/10.1145/2714064.2660234}
\showDOI{\tempurl}


\bibitem[Song and Lu(2017)]%
        {Song2017performance}
\bibfield{author}{\bibinfo{person}{Linhai Song} {and} \bibinfo{person}{Shan
  Lu}.} \bibinfo{year}{2017}\natexlab{}.
\newblock \showarticletitle{Performance diagnosis for inefficient loops}. In
  \bibinfo{booktitle}{\emph{Proceedings of the 39th International Conference on
  Software Engineering}} (Buenos Aires, Argentina) \emph{(\bibinfo{series}{ICSE
  '17})}. \bibinfo{publisher}{IEEE Press}, \bibinfo{pages}{370–380}.
\newblock
\showISBNx{9781538638682}
\urldef\tempurl%
\url{https://doi.org/10.1109/ICSE.2017.41}
\showDOI{\tempurl}


\bibitem[Su et~al\mbox{.}(2024)]%
        {su2024codeartbettercodemodels}
\bibfield{author}{\bibinfo{person}{Zian Su}, \bibinfo{person}{Xiangzhe Xu},
  \bibinfo{person}{Ziyang Huang}, \bibinfo{person}{Zhuo Zhang},
  \bibinfo{person}{Yapeng Ye}, \bibinfo{person}{Jianjun Huang}, {and}
  \bibinfo{person}{Xiangyu Zhang}.} \bibinfo{year}{2024}\natexlab{}.
\newblock \bibinfo{title}{CodeArt: Better Code Models by Attention
  Regularization When Symbols Are Lacking}.
\newblock
\newblock
\showeprint[arxiv]{2402.11842}~[cs.SE]
\urldef\tempurl%
\url{https://arxiv.org/abs/2402.11842}
\showURL{%
\tempurl}


\bibitem[Szebenyi et~al\mbox{.}(2009)]%
        {Szebenyi2009space}
\bibfield{author}{\bibinfo{person}{Zolt\'{a}n Szebenyi}, \bibinfo{person}{Felix
  Wolf}, {and} \bibinfo{person}{Brian J.~N. Wylie}.}
  \bibinfo{year}{2009}\natexlab{}.
\newblock \showarticletitle{Space-efficient time-series call-path profiling of
  parallel applications}. In \bibinfo{booktitle}{\emph{Proceedings of the
  Conference on High Performance Computing Networking, Storage and Analysis}}
  (Portland, Oregon) \emph{(\bibinfo{series}{SC '09})}.
  \bibinfo{publisher}{Association for Computing Machinery},
  \bibinfo{address}{New York, NY, USA}, Article \bibinfo{articleno}{37},
  \bibinfo{numpages}{12}~pages.
\newblock
\showISBNx{9781605587448}
\urldef\tempurl%
\url{https://doi.org/10.1145/1654059.1654097}
\showDOI{\tempurl}


\bibitem[Tan et~al\mbox{.}(2024)]%
        {tan2024llm4decompiledecompilingbinarycode}
\bibfield{author}{\bibinfo{person}{Hanzhuo Tan}, \bibinfo{person}{Qi Luo},
  \bibinfo{person}{Jing Li}, {and} \bibinfo{person}{Yuqun Zhang}.}
  \bibinfo{year}{2024}\natexlab{}.
\newblock \bibinfo{title}{LLM4Decompile: Decompiling Binary Code with Large
  Language Models}.
\newblock
\newblock
\showeprint[arxiv]{2403.05286}~[cs.PL]
\urldef\tempurl%
\url{https://arxiv.org/abs/2403.05286}
\showURL{%
\tempurl}


\bibitem[Vilk and Berger(2018)]%
        {Vilk2018Bleak}
\bibfield{author}{\bibinfo{person}{John Vilk} {and} \bibinfo{person}{Emery~D.
  Berger}.} \bibinfo{year}{2018}\natexlab{}.
\newblock \showarticletitle{BLeak: automatically debugging memory leaks in web
  applications}.
\newblock \bibinfo{journal}{\emph{SIGPLAN Not.}} \bibinfo{volume}{53},
  \bibinfo{number}{4} (\bibinfo{date}{June} \bibinfo{year}{2018}),
  \bibinfo{pages}{15–29}.
\newblock
\showISSN{0362-1340}
\urldef\tempurl%
\url{https://doi.org/10.1145/3296979.3192376}
\showDOI{\tempurl}


\bibitem[Wei et~al\mbox{.}(2024)]%
        {wei2024chainofthought}
\bibfield{author}{\bibinfo{person}{Jason Wei}, \bibinfo{person}{Xuezhi Wang},
  \bibinfo{person}{Dale Schuurmans}, \bibinfo{person}{Maarten Bosma},
  \bibinfo{person}{Brian Ichter}, \bibinfo{person}{Fei Xia},
  \bibinfo{person}{Ed~H. Chi}, \bibinfo{person}{Quoc~V. Le}, {and}
  \bibinfo{person}{Denny Zhou}.} \bibinfo{year}{2024}\natexlab{}.
\newblock \showarticletitle{Chain-of-thought prompting elicits reasoning in
  large language models}. In \bibinfo{booktitle}{\emph{Proceedings of the 36th
  International Conference on Neural Information Processing Systems}} (New
  Orleans, LA, USA) \emph{(\bibinfo{series}{NIPS '22})}.
  \bibinfo{publisher}{Curran Associates Inc.}, \bibinfo{address}{Red Hook, NY,
  USA}, Article \bibinfo{articleno}{1800}, \bibinfo{numpages}{14}~pages.
\newblock
\showISBNx{9781713871088}


\bibitem[Weng et~al\mbox{.}(2023)]%
        {Weng2023vProf}
\bibfield{author}{\bibinfo{person}{Lingmei Weng}, \bibinfo{person}{Yigong Hu},
  \bibinfo{person}{Peng Huang}, \bibinfo{person}{Jason Nieh}, {and}
  \bibinfo{person}{Junfeng Yang}.} \bibinfo{year}{2023}\natexlab{}.
\newblock \showarticletitle{Effective Performance Issue Diagnosis with
  Value-Assisted Cost Profiling}. In \bibinfo{booktitle}{\emph{Proceedings of
  the Eighteenth European Conference on Computer Systems}} (Rome, Italy)
  \emph{(\bibinfo{series}{EuroSys '23})}. \bibinfo{publisher}{Association for
  Computing Machinery}, \bibinfo{address}{New York, NY, USA},
  \bibinfo{pages}{1–17}.
\newblock
\showISBNx{9781450394871}
\urldef\tempurl%
\url{https://doi.org/10.1145/3552326.3587444}
\showDOI{\tempurl}


\bibitem[Weng et~al\mbox{.}(2021)]%
        {Lingmei2021Argus}
\bibfield{author}{\bibinfo{person}{Lingmei Weng}, \bibinfo{person}{Peng Huang},
  \bibinfo{person}{Jason Nieh}, {and} \bibinfo{person}{Junfeng Yang}.}
  \bibinfo{year}{2021}\natexlab{}.
\newblock \showarticletitle{Argus: Debugging Performance Issues in Modern
  Desktop Applications with Annotated Causal Tracing}. In
  \bibinfo{booktitle}{\emph{2021 USENIX Annual Technical Conference (USENIX ATC
  21)}}. \bibinfo{publisher}{USENIX Association}, \bibinfo{pages}{193--207}.
\newblock
\showISBNx{978-1-939133-23-6}
\urldef\tempurl%
\url{https://www.usenix.org/conference/atc21/presentation/weng}
\showURL{%
\tempurl}


\bibitem[Xia et~al\mbox{.}(2023)]%
        {Xia2023APR}
\bibfield{author}{\bibinfo{person}{Chunqiu~Steven Xia},
  \bibinfo{person}{Yuxiang Wei}, {and} \bibinfo{person}{Lingming Zhang}.}
  \bibinfo{year}{2023}\natexlab{}.
\newblock \showarticletitle{Automated Program Repair in the Era of Large
  Pre-Trained Language Models}. In \bibinfo{booktitle}{\emph{Proceedings of the
  45th International Conference on Software Engineering}} (Melbourne, Victoria,
  Australia) \emph{(\bibinfo{series}{ICSE '23})}. \bibinfo{publisher}{IEEE
  Press}, \bibinfo{pages}{1482–1494}.
\newblock
\showISBNx{9781665457019}
\urldef\tempurl%
\url{https://doi.org/10.1109/ICSE48619.2023.00129}
\showDOI{\tempurl}


\bibitem[Xia and Zhang(2024)]%
        {Xia2024ChatRepair}
\bibfield{author}{\bibinfo{person}{Chunqiu~Steven Xia} {and}
  \bibinfo{person}{Lingming Zhang}.} \bibinfo{year}{2024}\natexlab{}.
\newblock \showarticletitle{Automated Program Repair via Conversation: Fixing
  162 out of 337 Bugs for \$0.42 Each using ChatGPT}. In
  \bibinfo{booktitle}{\emph{Proceedings of the 33rd ACM SIGSOFT International
  Symposium on Software Testing and Analysis}} (Vienna, Austria)
  \emph{(\bibinfo{series}{ISSTA 2024})}. \bibinfo{publisher}{Association for
  Computing Machinery}, \bibinfo{address}{New York, NY, USA},
  \bibinfo{pages}{819–831}.
\newblock
\showISBNx{9798400706127}
\urldef\tempurl%
\url{https://doi.org/10.1145/3650212.3680323}
\showDOI{\tempurl}


\bibitem[Xiao et~al\mbox{.}(2013)]%
        {Xiao2013delta}
\bibfield{author}{\bibinfo{person}{Xusheng Xiao}, \bibinfo{person}{Shi Han},
  \bibinfo{person}{Dongmei Zhang}, {and} \bibinfo{person}{Tao Xie}.}
  \bibinfo{year}{2013}\natexlab{}.
\newblock \showarticletitle{Context-sensitive delta inference for identifying
  workload-dependent performance bottlenecks}. In
  \bibinfo{booktitle}{\emph{Proceedings of the 2013 International Symposium on
  Software Testing and Analysis}} (Lugano, Switzerland)
  \emph{(\bibinfo{series}{ISSTA 2013})}. \bibinfo{publisher}{Association for
  Computing Machinery}, \bibinfo{address}{New York, NY, USA},
  \bibinfo{pages}{90–100}.
\newblock
\showISBNx{9781450321594}
\urldef\tempurl%
\url{https://doi.org/10.1145/2483760.2483784}
\showDOI{\tempurl}


\bibitem[Xie et~al\mbox{.}(2024)]%
        {xie2024resym}
\bibfield{author}{\bibinfo{person}{Danning Xie}, \bibinfo{person}{Zhuo Zhang},
  \bibinfo{person}{Nan Jiang}, \bibinfo{person}{Xiangzhe Xu},
  \bibinfo{person}{Lin Tan}, {and} \bibinfo{person}{Xiangyu Zhang}.}
  \bibinfo{year}{2024}\natexlab{}.
\newblock \showarticletitle{ReSym: Harnessing LLMs to Recover Variable and Data
  Structure Symbols from Stripped Binaries}. In
  \bibinfo{booktitle}{\emph{Proceedings of the 2024 ACM SIGSAC Conference on
  Computer and Communications Security}}.
\newblock


\bibitem[Yang et~al\mbox{.}(2024)]%
        {Yang2024CREF}
\bibfield{author}{\bibinfo{person}{Boyang Yang}, \bibinfo{person}{Haoye Tian},
  \bibinfo{person}{Weiguo Pian}, \bibinfo{person}{Haoran Yu},
  \bibinfo{person}{Haitao Wang}, \bibinfo{person}{Jacques Klein},
  \bibinfo{person}{Tegawend\'{e}~F. Bissyand\'{e}}, {and}
  \bibinfo{person}{Shunfu Jin}.} \bibinfo{year}{2024}\natexlab{}.
\newblock \showarticletitle{CREF: An LLM-Based Conversational Software Repair
  Framework for Programming Tutors}. In \bibinfo{booktitle}{\emph{Proceedings
  of the 33rd ACM SIGSOFT International Symposium on Software Testing and
  Analysis}} (Vienna, Austria) \emph{(\bibinfo{series}{ISSTA 2024})}.
  \bibinfo{publisher}{Association for Computing Machinery},
  \bibinfo{address}{New York, NY, USA}, \bibinfo{pages}{882–894}.
\newblock
\showISBNx{9798400706127}
\urldef\tempurl%
\url{https://doi.org/10.1145/3650212.3680328}
\showDOI{\tempurl}


\bibitem[Yhuelf(2021)]%
        {yhuelf_pg_stat_statements}
\bibfield{author}{\bibinfo{person}{Yhuelf}.} \bibinfo{year}{2021}\natexlab{}.
\newblock \bibinfo{title}{Diagnosing Bottlenecks with pg\_stat\_statements}.
\newblock
\newblock
\urldef\tempurl%
\url{https://yhuelf.github.io/2021/09/30/pg_stat_statements_bottleneck.html}
\showURL{%
\tempurl}


\bibitem[Yuan et~al\mbox{.}(2010)]%
        {Yuan2010SherLog}
\bibfield{author}{\bibinfo{person}{Ding Yuan}, \bibinfo{person}{Haohui Mai},
  \bibinfo{person}{Weiwei Xiong}, \bibinfo{person}{Lin Tan},
  \bibinfo{person}{Yuanyuan Zhou}, {and} \bibinfo{person}{Shankar Pasupathy}.}
  \bibinfo{year}{2010}\natexlab{}.
\newblock \showarticletitle{SherLog: error diagnosis by connecting clues from
  run-time logs}. In \bibinfo{booktitle}{\emph{Proceedings of the Fifteenth
  International Conference on Architectural Support for Programming Languages
  and Operating Systems}} (Pittsburgh, Pennsylvania, USA)
  \emph{(\bibinfo{series}{ASPLOS XV})}. \bibinfo{publisher}{Association for
  Computing Machinery}, \bibinfo{address}{New York, NY, USA},
  \bibinfo{pages}{143–154}.
\newblock
\showISBNx{9781605588391}
\urldef\tempurl%
\url{https://doi.org/10.1145/1736020.1736038}
\showDOI{\tempurl}


\bibitem[Zhang et~al\mbox{.}(2017)]%
        {Zhang2017Pensieve}
\bibfield{author}{\bibinfo{person}{Yongle Zhang}, \bibinfo{person}{Serguei
  Makarov}, \bibinfo{person}{Xiang Ren}, \bibinfo{person}{David Lion}, {and}
  \bibinfo{person}{Ding Yuan}.} \bibinfo{year}{2017}\natexlab{}.
\newblock \showarticletitle{Pensieve: Non-Intrusive Failure Reproduction for
  Distributed Systems using the Event Chaining Approach}. In
  \bibinfo{booktitle}{\emph{Proceedings of the 26th Symposium on Operating
  Systems Principles}} (Shanghai, China) \emph{(\bibinfo{series}{SOSP '17})}.
  \bibinfo{publisher}{Association for Computing Machinery},
  \bibinfo{address}{New York, NY, USA}, \bibinfo{pages}{19–33}.
\newblock
\showISBNx{9781450350853}
\urldef\tempurl%
\url{https://doi.org/10.1145/3132747.3132768}
\showDOI{\tempurl}


\bibitem[Zhou et~al\mbox{.}(2018)]%
        {yang2016wperf}
\bibfield{author}{\bibinfo{person}{Fang Zhou}, \bibinfo{person}{Yifan Gan},
  \bibinfo{person}{Sixiang Ma}, {and} \bibinfo{person}{Yang Wang}.}
  \bibinfo{year}{2018}\natexlab{}.
\newblock \showarticletitle{{wPerf}: Generic {Off-CPU} Analysis to Identify
  Bottleneck Waiting Events}. In \bibinfo{booktitle}{\emph{13th USENIX
  Symposium on Operating Systems Design and Implementation (OSDI 18)}}.
  \bibinfo{publisher}{USENIX Association}, \bibinfo{address}{Carlsbad, CA},
  \bibinfo{pages}{527--543}.
\newblock
\showISBNx{978-1-939133-08-3}
\urldef\tempurl%
\url{https://www.usenix.org/conference/osdi18/presentation/zhou}
\showURL{%
\tempurl}


\end{thebibliography}
